\documentclass[pra,twocolumn,showpacs,amsmath,amssymb,superscriptaddress]{revtex4-2}
\usepackage[utf8]{inputenc}
\usepackage[T1]{fontenc}
\usepackage{graphicx}
\usepackage{grffile}
\usepackage{longtable}
\usepackage{wrapfig}
\usepackage{rotating}
\usepackage[normalem]{ulem}
\usepackage{amsmath}
\usepackage{textcomp}
\usepackage{amssymb}
\usepackage{capt-of}
\usepackage{hyperref}
\usepackage{parskip}
\usepackage{mathrsfs}
\usepackage[margin= 0.75in]{geometry}
\usepackage[braket, qm]{qcircuit}
\usepackage[english]{babel}
\usepackage{times}
\usepackage{bbm}
\usepackage{etoolbox}
\usepackage{makecell} % force line break in table cell
\usepackage{tcolorbox} % if you want to make a fancy box around some equations
\usepackage{fleqn} % automatically adds () around equation refs.
\usepackage{tikz}
\usepackage{amsthm}
\usepackage{amssymb}
\usepackage{chngcntr}

\usepackage[capitalise]{cleveref}

\newcommand{\RNum}[1]{\uppercase\expandafter{\romannumeral #1\relax}}

\newcommand{\bes} {\begin{subequations}}
\newcommand{\ees} {\end{subequations}}
\def\>{\rangle}
\def\<{\langle}
\def\Tr{\mathrm{Tr}}
\def\s{\sigma}
\def\d{\delta}
\newcommand{\ketb}[2]{|{#1}\>\!\<#2|}
\newcommand{\mc}[1]{\mathcal{#1}}
\newcommand{\ba}{\begin{align}}
\newcommand{\ea}{\end{align}}
\newcommand{\beq}{\begin{equation}}
\newcommand{\eeq}{\end{equation}}

\date{}
\hypersetup{
 pdfauthor={Haimeng Zhang (haimeng@usc.edu)},
 pdftitle={Noise characterization and suppression in superconducting qubits},
 pdflang={English}, colorlinks=true,  linkcolor=cyan, citecolor=magenta, filecolor=magenta, urlcolor=cyan, runcolor=cyan } 
\begin{document}
\title{Predicting non-Markovian superconducting qubit dynamics from tomographic reconstruction}

\author{Haimeng Zhang}
\affiliation{Department of Electrical \& Computer Engineering, University of Southern California,
Los Angeles, CA 90089, USA}
\affiliation{Center for Quantum Information Science \& Technology, University of
Southern California, Los Angeles, California 90089, USA}

\author{Bibek Pokharel}
\affiliation{Center for Quantum Information Science \& Technology, University of
Southern California, Los Angeles, California 90089, USA}
\affiliation{Department of Physics \& Astronomy, University of Southern California,
Los Angeles, California 90089, USA}

\author{E. M. Levenson-Falk}
\affiliation{Center for Quantum Information Science \& Technology, University of
Southern California, Los Angeles, California 90089, USA}
\affiliation{Department of Physics \& Astronomy, University of Southern California,
Los Angeles, California 90089, USA}

\author{Daniel Lidar}
\affiliation{Department of Electrical \& Computer Engineering, University of Southern California,
Los Angeles, CA 90089, USA}
\affiliation{Center for Quantum Information Science \& Technology, University of
Southern California, Los Angeles, California 90089, USA}
\affiliation{Department of Physics \& Astronomy, University of Southern California,
Los Angeles, California 90089, USA}
\affiliation{Department of Chemistry, University of Southern California, Los Angeles,
CA 90089, USA}

\date{\today} % Leave empty to omit a date

\begin{abstract}
Non-Markovian noise presents a particularly relevant challenge in understanding and combating decoherence in quantum computers, yet is challenging to capture in terms of simple models. Here we show that a simple phenomenological dynamical model known as the post-Markovian master equation (PMME) accurately captures and predicts non-Markovian noise in a superconducting qubit system. The PMME is constructed using experimentally measured state dynamics of an IBM Quantum Experience cloud-based quantum processor, and the model thus constructed successfully predicts the non-Markovian dynamics observed in later experiments. The model also allows the extraction of information about cross-talk and measures of non-Markovianity. We demonstrate definitively that the PMME model predicts subsequent dynamics of the processor better than the standard Markovian master equation.
\end{abstract}
\maketitle
\section{Introduction}

A central challenge in controlling and programming quantum processors is to overcome noise. 
Open quantum system dynamics are often modeled using the Gorini-Kossakowski-Sudarshan-Lindblad (GKSL) master equation~\cite{gorini_completely_1976,Lindblad:76}, also commonly known as the Lindblad master equation (LME). The LME is completely positive and is formally easily solvable. However, the LME is derived under the assumption of Markovianity. Loosely, this assumption amounts to an environment that is `memoryless' and is only valid when the system is weakly coupled to a bath whose characteristic timescale is much shorter than that of the system dynamics~\cite{Breuer:book,rivas_open_2012}. Although the Markovian assumption allows for significant simplifications, it is only an approximation and in reality it is often desirable to account for non-Markovian effects~\cite{RevModPhys.88.021002}.
This is true in particular in the case of the dynamics of superconducting qubit systems~\cite{Burkard:2009vd,Malekakhlagh:2016vw}. For example, it has been observed on the IBM Quantum Experience (IBMQE) processors~\cite{IBMQE} that the fidelity of a gate operation is conditional on the gate operation that preceded it in a gate sequence~\cite{morris_non-markovian_2019}. This is an example of non-Markovian noise, which introduces temporally correlated errors. Non-Markovian effects may arise from, for example, spatially correlated noise that arises from nonlocal external pulse controls, coherent errors caused by residual Hamiltonian terms, or stochastic errors due to slow environmental fluctuations. Such correlations, as well as correlated errors on multiple qubits, have been shown to be a leading source of failure in achieving quantum error correction~\cite{landahl_fault-tolerant_2011,Fowler:2012ys,devitt_quantum_2013}, and also in other near term quantum applications~\cite{sarovar_detecting_2020}. In other words, dealing with non-Markovianity will be vital to achieving fault-tolerant quantum computation~\cite{PhysRevA.73.052311,Aharonov:05,ng:032318,Ng:2011dn}.

Unfortunately, most device characterization and validation methods do not fully capture non-Markovian effects, as these methods either implicitly or explicitly make the Markovian approximation. For instance, the standard $T_1$ and $T_2$ measurements that quantify qubit lifetime assume exponential decay of the excited state population or the qubit coherence. Similarly, randomized benchmarking and gate set tomography~\cite{mavadia_experimental_2018} consider circuits of varying length and assume that the fidelity of corresponding operations decay as circuits become longer. However, on real quantum processors, recent studies~\cite{rudinger_experimental_2021,tripathi2021suppression} have observed deviations of the qubit dynamics from the prediction of a purely Markovian treatment.

In this work we focus on noise processes that govern the free (undriven) evolution of a superconducting quantum system. Here, the non-Markovian effects can be both coherent, e.g., due to unintentional crosstalk with neighboring qubits, or incoherent, e.g., due to coupling to magnetic impurities~\cite{RevModPhys.86.361}. However, a complete first-principles model for the system-bath interaction (for example, the formally exact Nakajima-Zwanzig (NZ) master equation~\cite{zwanzig_ensemble_1960}) may be infeasible to construct or too hard to solve numerically~\cite{Breuer:book}.
There has been extensive work on developing a set of master equations that are both easily solvable and account for non-Markovian effects, e.g., the Gaussian collapse model~\cite{ferialdi_exact_2016}, quantum collisional models~\cite{budini_embedding_2013} and the time-convolutionless master equations~\cite{Breuer:book}. Here, we choose to focus on the post-Markovian master equation (PMME)~\cite{shabani_completely_2005}, which includes bath memory effects via a phenomenological memory kernel $k(t)$ (see also~\cite{Lidar:2019aa} for an updated derivation). We choose the PMME for its conceptual and computational simplicity and because it has a closed form analytical solution in terms of a Laplace transform. It naturally interpolates between the exact dynamics (a completely positive map~\cite{Kraus:71}) and the Markovian Lindblad equation, and at the same time, retains complete positivity with an appropriate choice of the form and parameters for $k(t)$~\cite{shabani_completely_2005,chruscinski_conditions_2019,sutherland_non-markovianity_2018}, and remains analytically solvable.

The PMME we consider takes the following form: 
\begin{align}
\label{eq:pmme}
	\frac{d}{d t} \rho(t)&=\mathcal{L}_{0} \rho(t)\\
	& \quad + \mathcal{L}_{1} \int_{0}^{t} d t^{\prime} k\left(t^{\prime}\right) \exp \left[\left(\mathcal{L}_{0}+\mathcal{L}_{1}\right) t^{\prime}\right] \rho\left(t-t^{\prime}\right)\ .\notag
\end{align}
Here $\rho(t)$ is the reduced system state and $\mathcal{L}_0$ and $\mathcal{L}_1$ are Markovian (super-)generators in Lindblad form that describe the dissipative dynamics, where $\mathcal{L}_0$ can have additional Hermitian (i.e., unitary evolution generating) components. Non-Markovian effects in the evolution under the Lindblad superoperator $\mathcal{L}_1$ are introduced via a phenomenological memory kernel $k(t)$ to assign weights to the previous ``history'' of the system state. 
We note that Eq.~\eqref{eq:pmme} differs in an important way from the original PMME~\cite{shabani_completely_2005}, in that the latter did not contain $\mathcal{L}_{0}$ inside the integral. The reason we do this here will become apparent below; in essence, this allows us more flexibility in partitioning the various terms in the Lindbladian.  

In our protocol, the PMME model is constructed by fitting to an ensemble of time-domain tomography measurements. Hence, we call our protocol \emph{PMME tomography}. 
We demonstrate PMME tomography on an IBMQE device and then use the  PMME model thus constructed to quantify the degree of non-Markovianity. Our procedure predicts the non-Markovian effects we observe in future measurements and can model the information backflow from the environment to the system on the device we have tested. Our protocol provides a robust estimation method for a continuous dynamical model beyond the commonly assumed Markovian approximation, paving the way to more accurate modeling of noisy intermediate-scale quantum (NISQ) devices.

The structure of this paper is as follows. In Sec.~\ref{sec:methods}, we start by constructing the closed system model for the qubit evolution, introduce the non-Markovian open system model by proposing a Lindbladian and different memory kernel terms for Eq.~\eqref{eq:pmme}, describe the data collection procedure of the state tomography experiments, and finally use the estimated qubit states sampled during the evolution to find the best-fit PMME model parameters. In Sec.~\ref{sec:experiment} we apply the PMME model construction protocol to an IBMQE processor, discuss the descriptive power of the PMME models on the fitting data set, and discuss the models' predictive power by using them to predict qubit dynamics in a previously unseen testing data set. We also quantify the degree of non-Markovianity based on the constructed PMME model. In Sec.~\ref{sec:conc} we contrast our method with previous work, and provide a discussion of the results and conclusions. The Appendix contains various additional technical details.

\section{Methods}
\label{sec:methods}

\subsection{Closed system model}\label{sec:closed_system}
We consider a single qubit described by the effective Hamiltonian
\begin{align} 
\label{eq:hamiltonian}
H = -\frac{1}{2}\omega_z \sigma_z\ .
\end{align}
which is written in the rotating frame of the qubit drive, where $\omega_z$ accounts for the detuning between the qubit frequency and the drive frequency. 
In practice, the drive frequency is typically set to be the qubit frequency. The latter is determined via a calibration procedure, typically carried out on a single qubit with the rest of the neighboring, ``spectator'' qubits all in their ground states. A shift in qubit frequency can lead to a non-zero detuning between the qubit frequency and the drive, thus a non-zero $\omega_z$ in the effective Hamiltonian. In addition,   
the sign and the magnitude of $\omega_z$ can change depending on the initial state of the spectator qubits due to the presence of an always-on $ZZ$ interaction which arises from unintended coupling between the  qubit and its neighbors~\cite{tripathi_suppression_2021}. 
For these reasons, we include a Hamiltonian term in our model.  

\subsection{Open system models and their physical motivation}
\label{sec:model_motivations}
Our task is to find a model that best describes a time series of state tomography observations and find the best-fit parameters of that model. In this case the model itself is represented by the functional form of the memory kernel $k(t)$, while the model parameters are the parameters of the kernel function. We consider a sequence of models in order of increasing model complexity: the Lindblad model $\mathcal{M}_0$ and the PMME models $\mathcal{M}_1$ and $\mathcal{M}_2$. In general, we denote the models by $\mathcal{M}_i(\theta)$ where $\theta$ is a list of $p_i$ free model parameters, with $p$ increasing monotonically with $i$.

We take as our simplest model ($\mathcal{M}_0$) the Lindblad master equation
\bes
\begin{align}\label{eq:lindblad}
	\frac{d \rho}{d t}&=\mathcal{L}(\rho)=\mathcal{L}_0(\rho)+\mathcal{L}_1(\rho)\\
	&=-i[H,\rho] + \sum_{k} \gamma_{k}\left(V_{k} \rho V_{k}^{\dagger}-\frac{1}{2}\left\{V_{k}^{\dagger} V_{k}, \rho\right\}\right).
\end{align} 
\ees
As is clear from Eq.~\eqref{eq:pmme}, this is equivalent to a PMME with a delta-function kernel, $k_0(t)=\delta(t)$. In the notation of Eq.~\eqref{eq:pmme}, we choose the 
first generator as:
\begin{align}
\mathcal{L}_0 = \mathcal{H}+\mathcal{L}_\mathrm{GAD}\ , 
\end{align} 
which has a Hermitian component $\mathcal{H} (\rho)=-i\left[H, \rho\right]$ and a generalized amplitude-damping Lindbladian $\mathcal{L}_\mathrm{GAD}$  
with the Lindblad operators 
$V_k\in \{\sigma_+,\sigma_-\}$. We choose the second generator as a pure dephasing Lindbladian: 
\begin{align}
\mathcal{L}_1(\rho) = \gamma_{z}\left(\sigma_z \rho \sigma_z-\rho\right)\ .
\end{align} 
With this choice, which is motivated by our experiments with the IBMQE devices, the population decay ($T_1$) is essentially Markovian, as it is dominated by the $\mathcal{L}_{0} \rho(t)$ term outside the integral in Eq.~\eqref{eq:pmme}.

\begin{figure}[t]
\includegraphics[width=.8\linewidth]{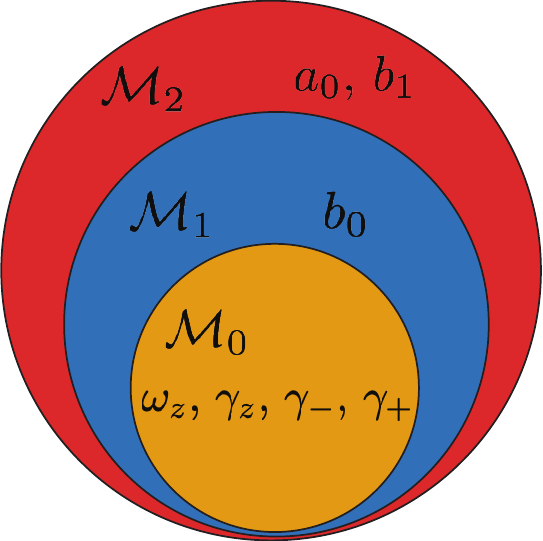}
\caption{The nested candidate models we use to describe the qubit free evolution: the Lindblad model $\mathcal{M}_0$, parameterized by a Hamiltonian term and the Lindbladian, and the PMME models $\mathcal{M}_{1},\mathcal{M}_2$ with their additional kernel parameters. 
}
\label{fig:1a}
\end{figure}

The parameters of interest to us are the following: $\omega_z$, the amplitude of the static $z$-field due to the always-on $ZZ$ coupling with the neighboring qubit(s); $\gamma_z$, the pure dephasing rate; $\Gamma_s=\gamma_++\gamma_-$ and $\Gamma_r=\gamma_+/\gamma_-$, respectively the sum and ratio of the excitation rate and the relaxation rate. The Kubo-Martin-Schwinger (KMS) condition~\cite{Breuer:book}, which states that the rate of excitation in a system is exponentially suppressed relative to the rate of relaxation at the same frequency [$\gamma(-\omega) = e^{-\beta\omega}\gamma(\omega)$, where $\beta$ is the inverse temperature and $\omega>0$], implies that $\Gamma_r<1$. This Lindblad model has a total of $p_0=4$ parameters.

To go beyond the Lindblad model $\mathcal{M}_0$, we consider two PMME models $\mathcal{M}_{1}$ and $\mathcal{M}_{2}$ with extra parameters for the memory kernel. We consider a family of kernels whose Laplace transform can be written as rational functions, i.e., 
\begin{align}
\tilde{k}_{1}(s)=\operatorname{Lap}[k(t)]=P(s)/Q(s)\ ,
\end{align} 
with polynomials $P(s)=a_{m}s^{m}+\dots+a_1s+ a_0$ and $Q(s)=b_{n}s^{n}+\dots+b_1s+ b_0$ of degree $m$ and $n$, respectively. This include a large class of kernels which can be expressed as linear combinations of functions of the form $t^de^{ct}$ for complex $c$ and integer $d$. For the PMME model $\mathcal{M}_1$ we choose the simplest kernel in this family: an exponentially decaying kernel 
\begin{align}\label{eq:kernel1}
k_1(t) = a_0 \exp(-b_0t)\  \Longleftrightarrow \ \tilde{k}_1(s)=a_0/(s+b_0)\ ,
\end{align} 
where henceforth we impose the constraint  $k(0)=1$ (since the normalization can always be absorbed into ${\mathcal{L}_1}$), which leads to $a_0=1$. This model thus has a total of $p_1=5$ free parameters: the Lindblad model parameters and $b_0$.

The more complex PMME model $\mathcal{M}_2$ has two additional free parameters $a_0$ and $b_1$: 
\begin{align}
\label{eq:kernel2}
k_{2}(t)&=\left\{\begin{array}{ll}
e^{-\frac{b_{1}}{2} t}\left[\frac{2 a_{0}-b_{1}}{2 \mu} \sinh (\mu t)+\cosh (\mu t)\right] & \text { if } B \geq 0 \\
e^{-\frac{b_{1}}{2} t}\left[\frac{2 a_{0}-b_{1}}{2 \mu} \sin (\mu t)+\cos (\mu t)\right] & \text { if } B <0 
\end{array}\right. \notag \\
& \Longleftrightarrow \tilde{k}_2(s)=(s+a_0)/(s^2+b_1s +b_0)\ ,
\end{align}
where $B=b_{1}^{2}-4 b_{0}$ and $\mu=\sqrt{|B|}$. 
 This model has a total of $p_2=7$ free parameters (the Lindblad parameters and $a_0$, $b_0$, and $b_1$). The sign of $B$ specifies whether the kernel is overdamped or underdamped. 
 
 As illustrated in 
Fig.~\ref{fig:1a}, the models $\mathcal{M}_0$, $\mathcal{M}_1$ and $\mathcal{M}_2$ are a sequence of nested models of increasing complexity; e.g., $\mathcal{M}_2$ reduces to $\mathcal{M}_1$ with the identifications $a_0=0, b_0=0$, and with a renaming of the parameter $b_1 \to b_0$ (since now the kernel function $\tilde{k}(s)$ has a lower degree).

The models $\{\mathcal{M}_i\}$ predict a functional form of the dynamics depending on the model variables ${\theta}$, so the predicted evolution of a state can be written as
\begin{align}
\rho^\mathrm{prd}(t) = f(t|\theta)\ ,
\end{align}
where $\theta=\{\omega_z,\gamma_z,\gamma_-,\gamma_+,\vec{a},\vec{b}\}$ is the list of model parameters.
The kernel parameters are $\vec{a}=\{a_0,\dots,a_{m-1}\}$, and $\vec{b}=\{b_0,\dots,b_{n-1}\}$, some of which may be constrained, in addition to the positivity constraint $\{\gamma_z,\gamma_-,\gamma_+\}>0$ and the KMS constraint $\gamma_+/\gamma_-<1$.

The goal of this procedure is to specify the master equation that governs the dynamics of the system. We formulate this in terms of the inverse problem: given a discrete time series of measurement records of the state, we seek the dynamical model $\mathcal{M}_i(\theta)$ that most closely matches the observations.

\subsection{Quantum State Tomography}
\label{sec:tomography}

\begin{figure}
\includegraphics[width=.8\linewidth]{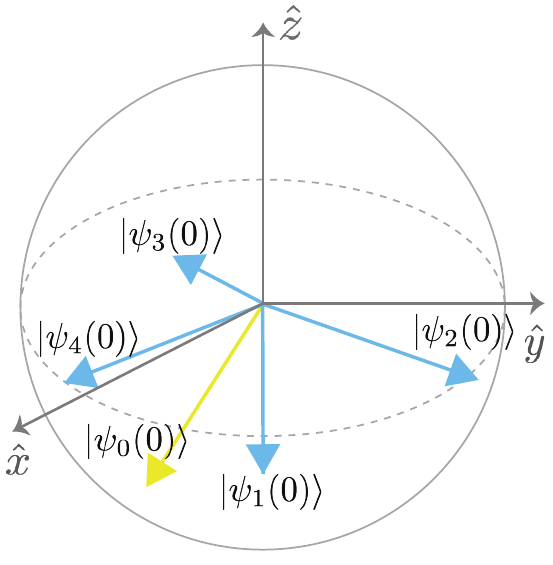}
\caption{The initial states used for the fitting data set (yellow) and the testing data set (blue). The initial states are chosen such that the states $|\psi_i(0)\rangle$, $i=0,1,2,3$ form a tetrahedron on the Bloch sphere, where $\ket{\psi_1(0)}=\ket{1}$ is the excited state. The state $|\psi_4(0)\rangle$ is a fixed, randomly chosen state on the Bloch sphere. The same set of five states are used in all our experiments.}
\label{fig:1b}
\end{figure}

\begin{table}[t]
\begin{center}
\begin{tabular}{ |c|c|}
\hline
 $|\psi_0(0)\rangle$ & $\left(\sqrt{\frac{8}{9}}, 0, -\frac{1}{3}\right)$  \\ 
 \hline
 $|\psi_1(0)\rangle$ &  $\left(0,0,-1\right)$ \\
 \hline
 $|\psi_2(0)\rangle$ & $\left(-\sqrt{\frac{2}{9}}, \sqrt{\frac{2}{3}}, -\frac{1}{3}\right)$  \\  
 \hline
 $|\psi_3(0)\rangle$ & $\left(-\sqrt{\frac{2}{9}},-\sqrt{\frac{2}{3}},-\frac{1}{3}\right)$  \\
 \hline
 $|\psi_4(0)\rangle$ &   $\left(0.50,-0.75,-0.41\right)$\\
\hline
\end{tabular}
\end{center}
\caption{The set of initial states $\mathbf{P}$ used in our experiments, corresponding to the states shown in Fig.~\ref{fig:1c}.}
\label{tab:1}
\end{table}

To get complete qubit state information, we perform state tomography on a single qubit of the  \textit{ibmq\_athens} processor (see Appendix~\ref{app:sysinfi}).
The data is collected with the main qubit state $\rho_S$ initialized in one of the five states in the preparation set $\mathbf{P}=\{\ket{\psi_i(0)}\}_{i=0}^{i=4}$; and the rest of the processor's spectator qubits initialized in the ground state $\ket{0}$. The states in $\mathbf{P}$ are illustrated in 
Fig.~\ref{fig:1c} and specified in Table~\ref{tab:1}.
After the qubit initialization, the main qubit undergoes free evolution for a variable time $t_j$, and then state tomography is performed to construct the density matrix, augmented by measurement error mitigation (see Appendix~\ref{app:MEM}). 
Specifically, the circuits of the tomography experiment contain the following steps:
\begin{enumerate}
	\item State preparation: the qubit is initialized in the ground state, and a state-preparation gate is applied to initialize the qubit in one of the five fiducial states $|\psi_i(0)\rangle$.
	\item Evolution: the qubit undergoes free evolution, with a variable evolution time. This corresponds to applying a sequence of identity gates $\mathbb{I}$ and sweeping the number of gates. 
	\item Measurement: one of the three single-qubit gates ($\mathbb{I}$, $\mathsf{M_X}$, or $\mathsf{M_Y}$)
is applied before the measurement, corresponding to measurement in the Pauli $z$, $x$, and $y$-bases respectively. Here $\mathsf{M_X}=H$ is a Hadamard gate and $\mathsf{M_Y}=HS^\dagger$ where $S$ is a phase gate. We record the measurement outcomes $0$ or $1$. 
\end{enumerate}
The steps above are repeated for all combinations of initial states, free evolution duration, and measurement basis, and each combination is repeated for $N_s=8192$ shots. Letting $N_{jk}$ denote the number of times out of $N_s$ shots that outcome 1 occurred at time $t_j$ in measurement basis $k$, the state tomography raw data are the recorded outcome counts,
\begin{align}
	\{N_{jk}|j=1,\dots,n_t,k\in\{x,y,z\}\}	
\end{align}
where $n_t$ is the total number of time points. We perform Bayesian measurement error mitigation on the raw data (see the Appendix for details) and the measurement mitigated data is then fed into a maximum likelihood estimation (MLE) routine. This routine determines 
the qubit state $\hat{\rho}(t_j)$ at time $t_j$, represented by the experimentally-measured Bloch vector $\vec{v}^\mathrm{exp}(t_j)$ as $\hat{\rho}(t_j)=\frac{1}{2}(\mathbb{I}+\vec{v}^\mathrm{exp}(t_j)\cdot\vec{\sigma})$. The uncertainties associated with the tomographically constructed states, i.e., the standard deviations of the corresponding Bloch vector components, denoted as $\sigma_{jk},\, k\in\{x,y,z\}$, are estimated by Bayesian bootstrapping. The collected data sets are divided into the model fitting set $\{\hat{\rho}\}_\mathrm{fit}$ and the model testing set $\{\hat{\rho}\}_\mathrm{test}$. The former contains a time series of qubit dynamics with a single 
initial state $\ket{\psi_0(0)}$, the latter contains the other four time series with four different initial states $\{\ket{\psi_i(0)}\}_{i=1}^{i=4}$. 
Finally, we use the constructed qubit states $\{\hat{\rho}(t_j)\}_{j=0}^{j=n_t-1}$ from the fitting data set to construct the PMME description of the free-evolution channel.

\subsection{Model fitting}\label{sec:model_fitting}
Given the observations from the state tomography experiment, our task is to find the best-fit model parameters $\theta$, as parametrized in Sec.~\ref{sec:model_motivations}, of the dynamical model $\mathcal{M}_i(\theta)$.
To tackle the problem, we perform a maximum likelihood estimation (MLE), a well-studied method of determining unknown parameters of a model from a set of data. The input data of this MLE procedure is a time series of the qubit evolution $\{t_j,\hat{\rho}^\mathrm{exp}(t_j)\}_{j=0}^{j=N_t-1}$, initialized in one of the states in the set $\mathbf{P}$.  
The qubit state $\hat{\rho}^\mathrm{exp}$ at time $t_j$ is constructed from tomography experiment outcomes via a Qiskit state tomography MLE routine~\cite{fisher_quantum_nodate}.
In the fitting procedure, we seek to minimize the distance between the observed state $\hat{\rho}^\mathrm{exp}(t_j)$ and the model predicted state $\rho^\mathrm{prd}(t_j)$ for all the sampled time instances $t_j$. We define the following standard objective function in the least squares form~\cite{Wolberg:2005vv}:
\begin{align}
\label{eq:obj}
    \chi^2(\theta) = \sum_{j=0}^{N_t-1}\sum_{k=x,y,z}\frac{[\hat{v}_{jk}^\mathrm{exp}(t_j)-v_{jk}^\mathrm{prd}(t_j;\theta)]^2}{\sigma_{jk}^2},
\end{align}
where $\vec{v}_{jk}^\mathrm{exp}$ and $\vec{v}_{jk}^\mathrm{prd}$ denote the $k$th Bloch vector component of the qubit state at time $t_j$ constructed from the experimental observations and predicted by the model, respectively. The latter is obtained by solving the model $\mathcal{M}_i(\theta)$ for $\rho^\mathrm{prd}(t)$. Since the PMME admits a closed-form analytical solution, the evaluation of the objective function Eq.~\eqref{eq:obj} is efficient to compute 
under the assumption that the noise associated with each data point follows a normal distribution. Minimizing the least-squares function $\chi^2(\theta)$ for the model parameters $\theta$ is equivalent to maximizing the likelihood of observing the data set given that the underlying model is true. 

We would like the dynamical model to generate a CPTP map (though this is not strictly necessary, as non-CP maps are also valid physical models; see, e.g., Ref.~\cite{Dominy:2016xy}). For this reason, we restricted the Lindblad rates $\{\gamma_z,\gamma_+,\gamma_-\}$ to be positive. For the PMME models with different choices of kernel functions, we derive the condition that guarantees CPTP dynamics in Appendix~\ref{app:cptp}. 
We find that the conditions $|\xi_4| < 1$ and $|\xi_2|=|\xi_3|<\frac{\sqrt{(\Gamma_r+\xi_4)(1+\Gamma_r \xi_4)}}{1+\Gamma_r} < 1$ are necessary and sufficient for complete positivity, where $\xi_i(t)=\operatorname{Lap}^{-1}\left[\frac{1}{s-\lambda_i^0-\lambda_i^1k(s-\lambda_i)}\right]$ (the inverse Laplace transform), and $\lambda_i^0$, $\lambda_i^1$, and $\lambda_i$ are the eigenvalues of the matrix representation of $\mc{L}_0$, $\mc{L}_1$, and $\mc{L}$, respectively, in the Pauli basis (see Appendix~\ref{app:A}).

The best-fit PMME parameters are found by solving the following minimization problem:
\begin{align}
\label{eq:minimize}
\text { minimize } & \chi^2\left(\omega_{z}, \gamma_{z}, \gamma_{+}, \gamma_{-},\vec{a},\vec{b}\right) \\
\text { subject to } &  \gamma_{z}, \gamma_{+}, \gamma_{-}>0,\,\Gamma_r=\gamma_{+}/\gamma_{-}<1, \, k(0)=1\ . \notag
\end{align}

After we fit the models by solving Eq.~\eqref{eq:minimize}, we compute the trace-norm distance 
\begin{align}
    \mathcal{D}(t_j) = \frac{1}{2}\|\hat{\rho}^\mathrm{exp}(t_j)-\rho^\mathrm{prd}(t_j)\|_1,
    \label{eq:D}
\end{align}
where $\|A\|_1 = \mathrm{Tr}[\sqrt{A^\dag A}]$ is the trace norm. $\mathcal{D}(t_j) $ quantifies the probability with which one can optimally discriminate the experimentally observed $\hat{\rho}(t_j)$ from the predicted state $\rho(t_j)$~\cite{nielsen2010quantum}, and thus a small $\mathcal{D}(t_j) $ indicates an accurate prediction.

\section{Results}
\label{sec:experiment}
\begin{figure}
\includegraphics[width=\linewidth]{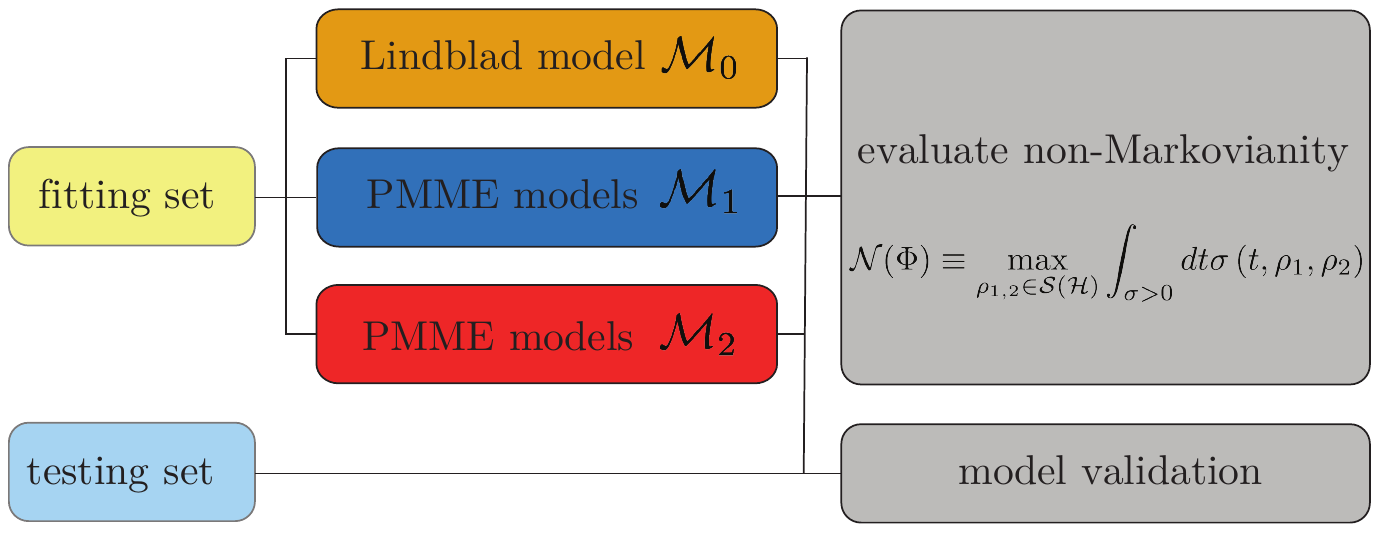}
\caption{Analysis protocol for PMME tomography. The tomography data sets for qubit free evolution with different initial states are divided into a fitting set (in our case with a single initial state $|\psi_0(0)\rangle$) and a testing set (in our case with initial states $|\psi_{1,2,3,4}(0)\rangle$). The fitting set is used to fit the dynamical models $\{\mathcal{M}_i\}$ using a classical optimizer based on the MLE method.}
\label{fig:1c}
\end{figure}

We test our approach for tomographic PMME construction in three increasingly challenging settings, summarized in 
Fig.~\ref{fig:1c}.
We start with time-series state tomography data for a single initial state on a transmon qubit (the fitting data set on $|\psi_0(0)\rangle$) and show that we can construct a sequence of nested PMME models to represent this evolution. We then compare the PMME models with the Markovian Lindblad counterpart and show that the PMME models provide a measurably better fit. Secondly, we test the PMME's predicted evolution for quantum states that were not used to construct the PMME model. Again, we show that the PMME model faithfully predicts the evolution even for these new states without requiring full process tomography, while the Lindblad model does not. Lastly, we compute the degree of non-Markovianity in the evolution using the tomographically constructed PMME model and show that it correctly approximates the observed non-Markovianity in the qubit free-evolution dynamics, which the Lindblad model fails to do.

\begin{figure*}[t]
  \centering
 \includegraphics[width=.9\linewidth]{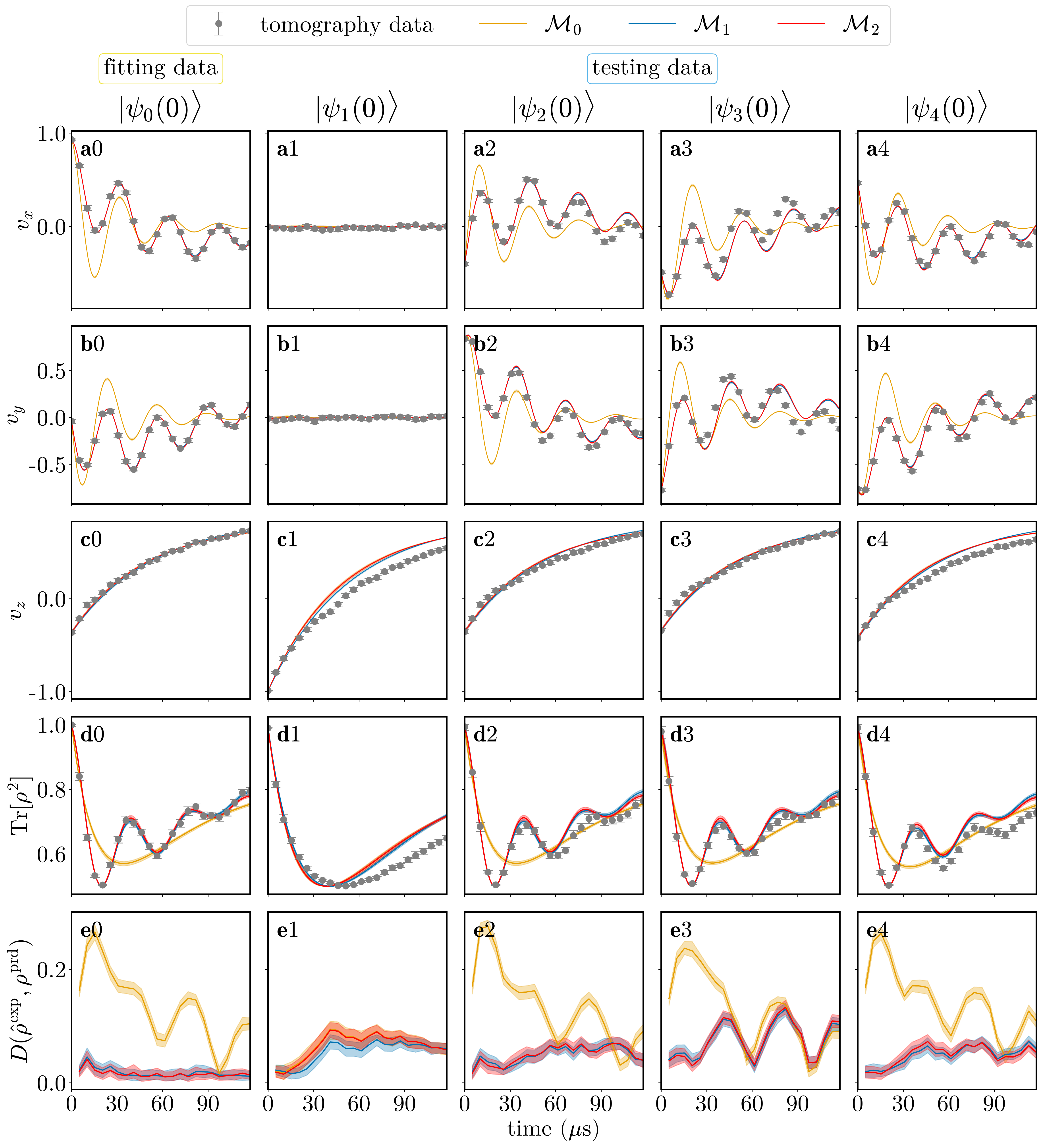}
  \caption{PMME tomography protocol applied to single-qubit free evolution, with the spectator qubits all in their ground state. (a0-c0) The free-evolution tomography data $\hat{v}^\mathrm{exp}(t)$ of the fitting data set with the qubit initialized in $|\psi_0(0)\rangle$ and the best-fit models for the Lindbladian model $\mathcal{M}_0$ (orange lines), the PMME model with kernel type 1, $\mathcal{M}_1$ (blue lines), and the PMME model with kernel type 2, $\mathcal{M}_2$ (red lines). The shaded regions denote the 95\% confidence region of the model predictions. (a1-c4) The free-evolution tomography data in the testing data set with the qubit initialized in $\{|\psi_{i}(0)\rangle\}_{i=1}^4$, and the prediction from the best-fit models from the fitting data set in (a0-c0). (d0-d4) The empirical purity $\Tr[\rho^{2}]$ of the qubit and that predicted by the best-fit models. (e0-e4) The distance between the tomographically constructed state and the state predicted by the best-fit models. All models perform equally well at predicting the dynamics of the excited state $|\psi_1\rangle=\ket{1}$ ($\mathcal{M}_{0}$ is obscured by $\mathcal{M}_{1}$ and $\mathcal{M}_{2}$), while the PMME models $\mathcal{M}_{1}$ and $\mathcal{M}_{2}$ predict the dynamics of $\{|\psi_{i}(0)\rangle\}_{i=2}^4$ better than the Lindbladian model. 
  }
    \label{fig:fit_and_test}
\end{figure*}

\subsection{Fitting set}\label{sec:fitting_set}

\begin{figure}
\includegraphics[width=\linewidth]{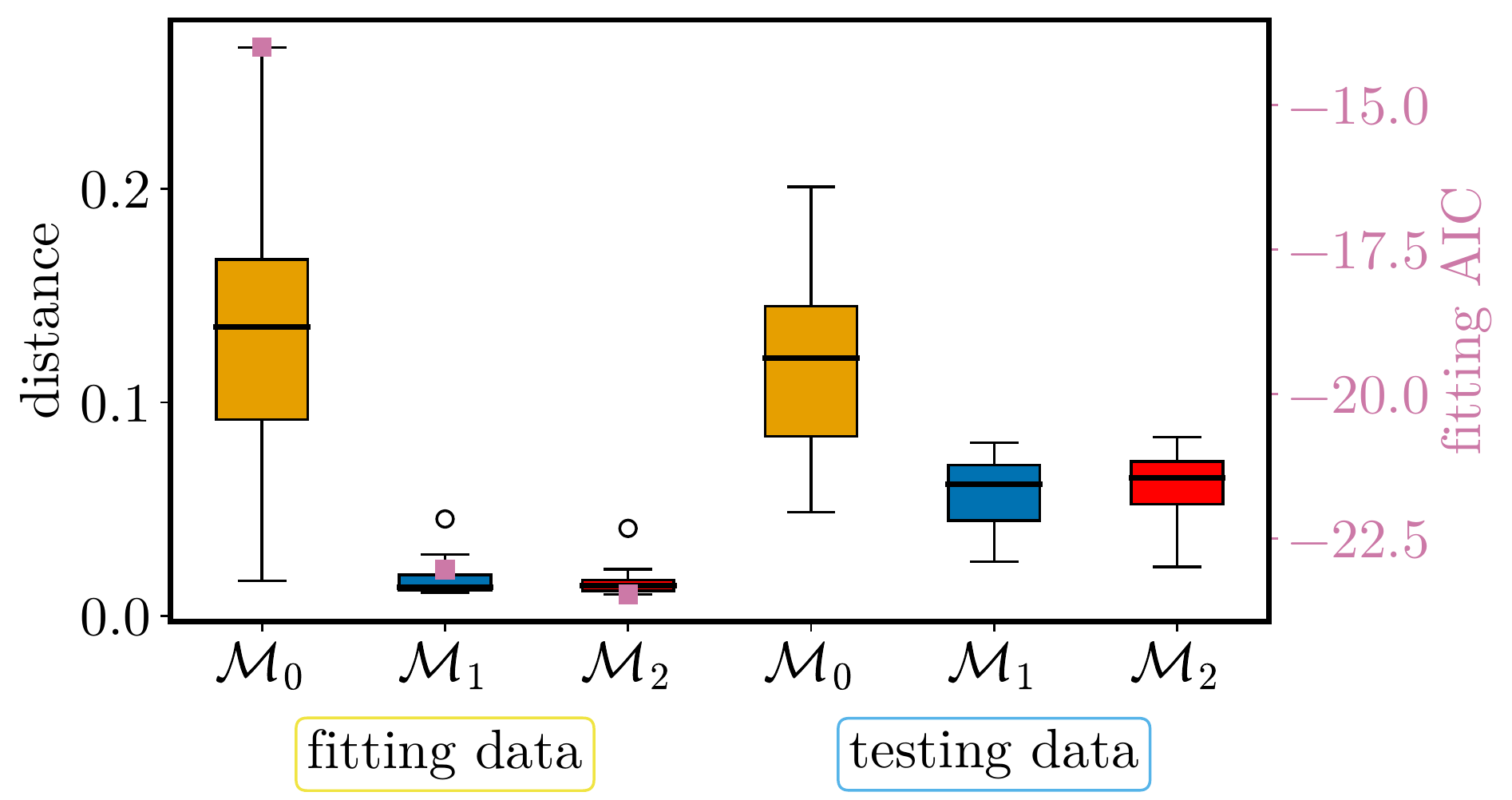}
\caption{The predictive ability of different models for the fitting data set (left) and the testing data set (right) for \textit{ibmq\_athens} data. For both datasets, we show the trace-norm distance between the tomographically constructed state $\{\hat{\rho}^\mathrm{exp}\}$ and the model predicted state $\{\rho^\mathrm{prd}\}$. The box plot shows the 5th, 50th (median), and 95th percentiles for this distance over $t_j$ and respective initial states.  The open circles denote extremal outliers. The PMME models $\mathcal{M}_{1}$ and $\mathcal{M}_{2}$ describe both the fitting data set and the testing data set better than the Lindblad model $\mathcal{M}_0$ (see Table~\ref{tab:2}). We also report the AIC of the models on the fitting data (purple squares, right axis), again with better performance by the non-Markovian models; note that all AIC values are negative.}
\label{fig:1d}
\end{figure}

First, we construct the dynamical models $\mathcal{M}_0$, $\mathcal{M}_1$ and $\mathcal{M}_2$ using the time-series tomography data of a single initial state. Our results are displayed in the first column of Fig.~\ref{fig:fit_and_test}, which shows the observed evolution $\vec{v}^\mathrm{exp}$ (grey circles with error bars) and the predicted evolution $\vec{v}^\mathrm{prd}$ (solid lines) from the constructed dynamical models. We plot the Bloch vector components in rows a, b, and c, the state purity $\mathrm{Tr}[\rho^2]$ in row d, and the trace-norm distance between the measured and the predicted state $ \mathcal{D}(t_j)$ [Eq.~\eqref{eq:D}] in row e.
The PMME models $\mathcal{M}_1$ and $\mathcal{M}_2$ both faithfully capture the evolution of the system. 
The AIC (Akaike information criterion)~\cite{akaike_new_1974} (a metric often used in model selection tasks, discussed in Appendix~\ref{app:D} in more detail) penalizes $\mathcal{M}_2$ for utilizing more parameters than $\mathcal{M}_1$ and so is slightly lower for $\mathcal{M}_1$ than for $\mathcal{M}_2$, as shown by the purple squares in 
Fig.~\ref{fig:1d}. In other words, $\mathcal{M}_1$ more than compensates for a slight loss of accuracy relative to $\mathcal{M}_2$ by reducing the number of fitting parameters. 

Both the Lindblad model $\mathcal{M}_0$ and the PMME models can account for a spurious longitudinal field component in the qubit Hamiltonian due to the always-on $ZZ$ interaction between neighboring qubits~\cite{ganzhorn_benchmarking_2020}. This spurious field effectively shifts the qubit frequency [Eq.~\eqref{eq:hamiltonian}] and manifests as the oscillations in the off-diagonal elements $v_x(t)$ and $v_y(t)$ of the density matrices of the qubit state (the first and second rows of Fig.~\ref{fig:fit_and_test}). However, a Hamiltonian term does not modify the purity $p=\Tr[\rho^2]$ since $\dot{p} = -i\Tr[\rho[H,\rho]]=0$, so purity oscillations must have a different origin. The fact that $\mathcal{M}_0$ is unable to capture the purity oscillations seen in the fourth row of Fig.~\ref{fig:fit_and_test}, while in contrast both $\mathcal{M}_1$ and $\mathcal{M}_2$ do display purity oscillations, is evidence of non-Markovianity, as we explore below in greater depth. The non-monotonic envelope of the purity oscillations is consistent with the non-unitality of the Lindbladian model~\cite{Lidar200682}.
Overall, in comparison to the Lindbladian model $\mathcal{M}_0$, the PMME-predicted evolution is significantly closer to the empirical data, as quantified by $\mathcal{D}(\hat{\rho}^\mathrm{exp},\rho^\mathrm{prd})$ (last row of Fig.~\ref{fig:fit_and_test}).

\begin{table}[t]
\begin{center}
\begin{tabular}{ |c|c|c|c|c|}
\hline
 & \multicolumn{2}{c|}{fitting data} & \multicolumn{2}{c|}{testing data}\\
 \hline
 & median & 95th percentile  & median  & 95th percentile  \\
 \hline
  $\mathcal{M}_0$ & 0.13 & 0.24 & 0.12 & 0.18 \\
 \hline
 $\mathcal{M}_1$ & 0.01 & 0.01 & 0.06 & 0.08 \\
 \hline
 $\mathcal{M}_2$ & 0.01 & 0.01 & 0.06 & 0.08 \\
 \hline
\end{tabular}
\end{center}
\caption{Trace-norm distances corresponding to the results shown in Fig.~\ref{fig:1d}.}
\label{tab:2}
\end{table}

\subsection{Testing set}
\label{sec:testing_set}
Next, we test how well the fitted models predict the evolution of states that were not used to fit these models. The last four columns of Fig.~\ref{fig:fit_and_test} represent four time series of quantum state evolutions in the testing data set with each of the different initial states $\{\ket{\psi_{1}(0)}\}_{i=1}^4$. The goal of this `testing set' is to validate whether the dynamical model is capable of describing the evolution for an \emph{arbitrary} single-qubit state. Rather than doing this by selecting random states, we chose a set of four states that is maximally separated on the Bloch sphere, i.e., four pure states on the vertices of a tetrahedron (see Fig.~\ref{fig:1b}). We emphasize that no fitting is done on this testing data; instead, we use the fits from the fitting data set to \emph{predict} the state evolution in the testing data set.

Once again, we find that the Lindbladian model's prediction provides a crude approximation to the evolution. In particular, for $| \psi_1(t) \rangle$, where the initial qubit state is the excited state, the amplitude damping process dominates, and the Markovian Lindblad model is sufficient to describe the dynamics. However, $| \psi_1(t) \rangle$ is an exception. For all other states, the PMME models are far more accurate, as is clear from Fig.~\ref{fig:fit_and_test} (a2-d4). This suggests that the non-Markovian effects mainly manifest in the evolution of the qubit phase coherence but not in the state populations, so the need for the more complex PMME models arises when dealing with states with coherence in the computational basis (this pointer basis -- the ground and excited states -- is einselected~\cite{Zurek:2003uu} due to thermal relaxation). Another observation visible from Fig.~\ref{fig:fit_and_test} (row d) is that the Markovian model's predictions become more accurate at relatively long evolution times, i.e., Markovian effects become more dominant on a timescale of $\sim 100\mu$s.

Overall, as shown in Fig.~\ref{fig:1d} and Table~\ref{tab:2}, the median and worst-case prediction distances of the PMME models $\mathcal{M}_1$ and $\mathcal{M}_2$ are very close, and substantially better than those of the Lindbladian model $\mathcal{M}_0$. In particular, consider the box plot in Fig.~\ref{fig:1d} reporting the statistics for the trace-norm distance across all sampling points $t_j$ and initial states of the fitting and testing datasets respectively. For the testing data, while the Lindblad model $\mathcal{M}_0$ has a median trace-norm distance of 0.12 and worst-case fidelity of 0.18, both the PMME models $\mathcal{M}_1$ and $\mathcal{M}_2$ have a median distance of 0.06 and the worst-case fidelity of 0.08.

Notably, the distance does increase slightly when going from the fitting data to the testing data, which can be attributed to the qubit's environment changing during the time between measurements; for example, the qubit relaxation time $T_1$ fluctuates~\cite{carroll_dynamics_2021}. Still, the PMME models provide a significantly closer correspondence to the fitting data set and more accurate predictions of the testing data sets than the Lindblad model does in both the average and worst cases. 
Figure~\ref{fig:fit_and_test} clearly shows that these differences in distance between $\mathcal{M}_0$ and $\mathcal{M}_{1,2}$ come from $\mathcal{M}_0$'s inability to capture oscillations about a non-zero mean in the Bloch vector components $v_x(t)$ and $v_y(t)$, which $\mathcal{M}_{1}$ and $\mathcal{M}_{2}$ do capture. Our results validate that while $\mathcal{M}_0$ is a good first Markovian approximation, the PMME models account for non-Markovian nuances. 

\subsection{Quantifying non-Markovianity}
\label{sec:non_mark}
Lastly, we test whether the PMME model can capture non-Markovianity during the qubit free evolution. To quantify this, we adopt the measure in Ref.~\cite{breuer_measure_2009} which uses the rate of change of the trace-norm distance between two quantum states under some noise channel $\Phi_t$:
\begin{align}\label{eq:derivative}
	\sigma\left(t, \rho_{1}, \rho_{2}\right) \equiv \frac{d}{d t} \mathcal{D}\left(\Phi_{t}\left(\rho_{1}(0)\right), \Phi_{t}\left(\rho_{2}(0)\right)\right)\ .
\end{align} 
Under Markovian dynamics [Eq.~\eqref{eq:lindblad}], the trace-norm distance between two quantum states is monotonically decreasing as a function of time, whereas non-Markovian dynamics violates this contractive property, i.e., there can be an increase in the trace-norm distance. In other words, non-Markovianity leads to revival of distinguishability between two states at some point during the evolution, and a process is non-Markovian if there exists any pair of initial states $\rho_{1}(0), \rho_{2}(0)$ and a time $t$ for which $\sigma\left(t, \rho_{1}, \rho_{2}\right)>0$~\cite{breuer_measure_2009}.

The measure for the degree of non-Markovianity of a quantum process is thus defined as:
\begin{align}\label{eq:blp}
	\mathcal{N}(\Phi) \equiv \max _{\rho_{1,2} \in \mathcal{S}(\mathcal{H})} \int_{\sigma>0} d t\, \sigma\left(t, \rho_{1}, \rho_{2}\right)	.
\end{align}
Here, the time-integration is over time intervals where $\sigma$ is positive, and the maximum is over all pairs of initial states. Thus, this quantity measures the total increase of distinguishability over the whole evolution time. We note that it is not normalized, and hence its actual numerical values are difficult to interpret. Its main significance is in the the fact that a non-zero non-Markovian measure means non-Markovian dynamics, and, as shown in Table~\ref{tab:doNM}, the experimentally obtained values are in fair agreement with the values predicted by our model. The non-Markovianity of the PMME is discussed in~\cite{sutherland_non-markovianity_2018} using this measure.

\begin{figure}
  \centering
  \includegraphics[width=\linewidth]{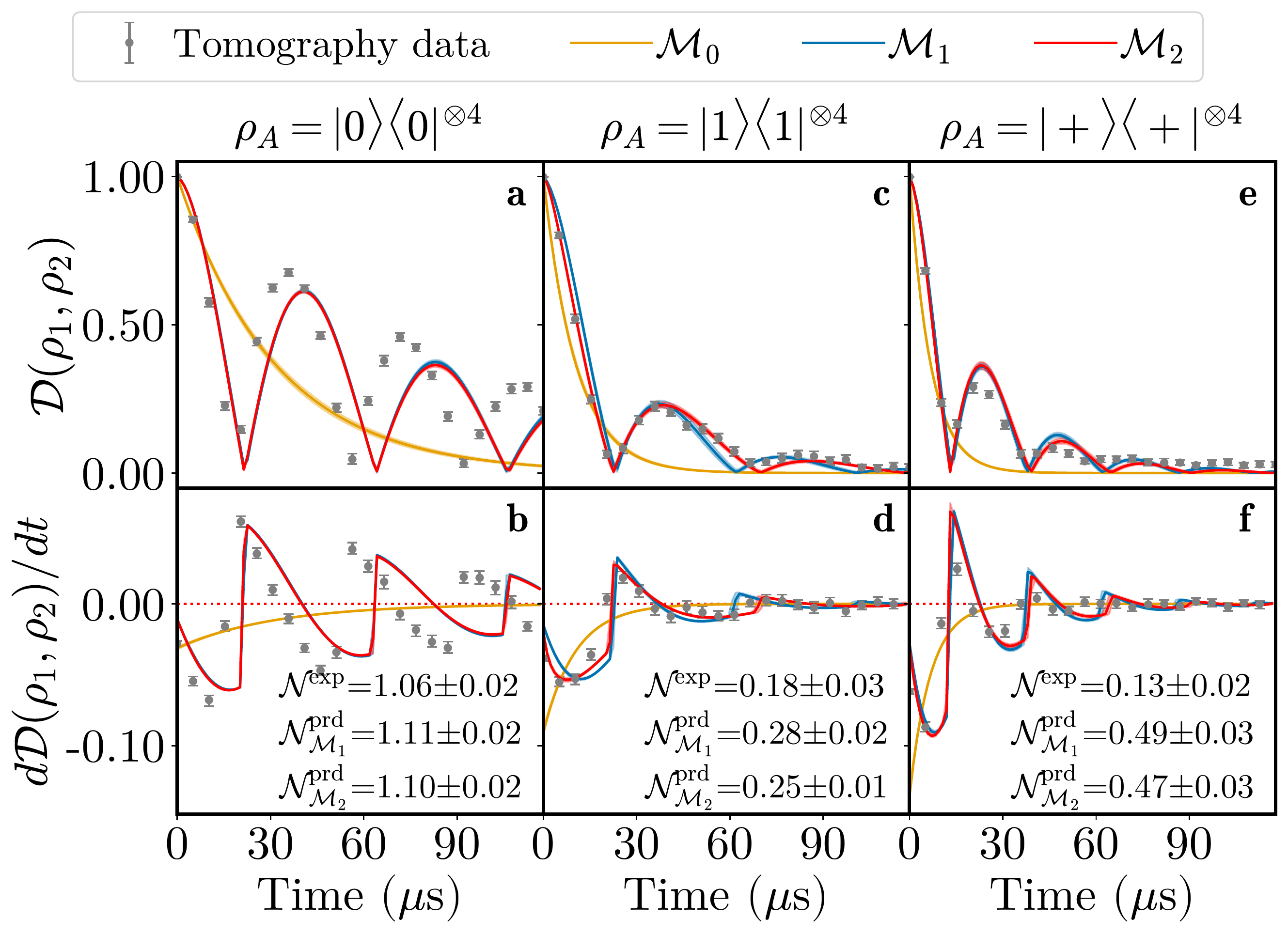}
  \caption{ 
  Non-Markovianity of qubit free-evolution dynamics for spectator qubits in the ground state (a,b), the excited state (c,d) and the $|+\rangle$ state (e,f). (a,c,e) The trace-norm distance $\mathcal{D}(\rho_1(t),\rho_2(t))$ predicted by the best-fit models (solid lines) and experimentally measured by performing free-evolution tomography with a pair of initial states $\rho_1(0)=|+\rangle\langle +|$ and $\rho_2(0)=|-\rangle\langle -|$ (grey circles). (b,d,f) The derivative $\sigma(t)$, defined in Eq.~\eqref{eq:derivative}, predicted by the best-fit models (solid lines), and approximated experimentally  using forward differencing based on the tomography data in (a,c,e) (grey circles).}
  \label{fig:nonMark}
\end{figure}

\begin{table}[t]
\begin{center}
\begin{tabular}{ |c|c|c|c|}
\hline
spec. qubit init. state & $\mathcal{N}^\mathrm{exp}$ & $\mathcal{N}^\mathrm{prd}_{\mathcal{M}_{1}}$  & $\mathcal{N}^\mathrm{prd}_{\mathcal{M}_{2}}$    \\
 \hline
 $\ket{0}^{\otimes 4}$ & $1.06 \pm 0.02$ & $1.11\pm 0.02$ & $1.10\pm 0.02$  \\
 \hline
 $\ket{1}^{\otimes 4}$  & $0.18 \pm 0.03$ & $0.28 \pm 0.02$ & $0.25\pm 0.01$ \\
 \hline
 $\ket{+}^{\otimes 4}$  & $0.13 \pm 0.02$ & $0.49\pm 0.03$ & $0.47\pm 0.03$\\
 \hline
\end{tabular}
\end{center}
\caption{Degree of non-Markovianity for three different initial states of the spectator qubits (column 1) as computed from the experimental data (column 2) and the two non-Markovian models (columns 3 and 4).}
\label{tab:doNM}
\end{table}

We performed state tomography on free evolution with the qubit initialized in the following two pairs of states: $\{\rho_1(0)=\ketb{+}{+},\rho_2(0)=\ketb{-}{-}\}$, and $\{\rho_1(0)=\ketb{+i}{+i},\rho_2(0)=\ketb{-i}{-i}\}$. From the time-series state tomography data, we see how the trace-norm distance $\mathcal{D}(\Phi_t(\rho_1(0)),\Phi_t(\rho_2(0)))$, starting from $\mathcal{D}=1$ (maximally distinguishable), evolves as a function of time, as plotted in the top row of Fig.~\ref{fig:nonMark}. Using forward differences to approximate the time derivative of our experimental data, we then approximate the quantity in Eq.~\eqref{eq:derivative}, plotted in the bottom row of Fig.~\ref{fig:nonMark}. By performing a linear interpolation on the discrete samples shown there, we further estimate the observed degree of non-Markovianity as defined in Eq.~\eqref{eq:blp} (but excluding the maximization it requires). The results we find for the degree of non-Markovianity are summarized in Table~\ref{tab:doNM}. 

Of course, $\mathcal{N}^\mathrm{prd}_{\mathcal{M}_0}\equiv 0$ for the Markovian Lindblad model. The experimental measure is estimated from the state tomography data,
while
the predicted non-Markovian measure
is calculated exactly from our dynamical models, since they provide a continuous description of the qubit state evolution. We train these models on a new fitting data set with the same initial state $\ket{\psi_0(0)}$ as the prior fitting data (Fig.~\ref{fig:1b}); the new fitting data was taken in the same batch of jobs as the trace-norm distance data in order to minimize the time delay between them, thus reducing the effect of systematic errors due to differences between batches. 
As seen in Table~\ref{tab:doNM}, the two PMME models converge to almost the same quantity in their predictions, despite the different forms of their kernels. The tomography experiments confirm that there is indeed an increase of distinguishability between quantum states during evolution. The predictions of the PMME models adequately match the observed quantity $\sigma(t_j)$, with some deviations which are due to numerical errors arising from using the finite difference formula to approximate Eq.~\eqref{eq:derivative}, and due to the run-to-run system fluctuations of the quantum processor~\cite{dasgupta_stability_2021}.

\section{Discussion}
\label{sec:conc}

We have developed and implemented a procedure to fit a family of phenomenological quantum master equations to time-series state tomography data.  We demonstrated this method by characterizing the free evolution of a single qubit on an IBMQE processor. From the constructed models, we conclude that the qubit Hamiltonian accounts for a residual longitudinal field due to the crosstalk with the neighboring qubits, and found the Lindblad rates that correspond to dephasing, spontaneous emission, and thermal excitation. However, a purely Markovian model provided a relatively poor fit to the data. We thus constructed post-Markovian master equation (PMME) models that contain a phenomenological bath memory kernel which account for the non-Markovianity of the dynamics -- something a Markovian Lindblad model cannot do. These PMME models provide a closer fit to the tomography data, and also much more accurately predict the future dynamics for different and new initial states that the models were not already fitted to.

Our PMME construction procedure is an alternative to other methods characterizing qubit noise processes such as process tomography \cite{Chuang:97c,boulant_robust_2003,Howard:06,MohseniLidar:06}, machine learning (ML) \cite{banchi_modelling_2018,krastanov_unboxing_2020,flurin_using_2020,luchnikov_machine_2020}, and shadow tomography~\cite{Aaronson-shadow-tomography,Huang:2020wo,levy2021classical}.
Compared to the first of these methods, our approach is less demanding in terms of both data collection and analysis, since it relies on state tomography, which generally (but not always \cite{MohseniLidar:06}) requires a number of measurements which is quadratically smaller in the Hilbert space dimension~\cite{Mohseni:2008ly}. The ML methods usually require a large training data set for the solution to converge since the ML model is usually overparametrized. In contrast, at least in the single-qubit setting, we have demonstrated that our method requires doing state tomography with only one initial state at multiple time points. Another advantage of our method is that the evaluation of the cost function in our problem is straightforward because the PMME is analytically solvable (see App.~\ref{app:B}). In contrast, the noise models that have been considered in ML so far are numerically challenging because they involve solving the Nakajima-Zwanzig master equation, whose complexity is proportional to the kernel's length~\cite{krastanov_unboxing_2020}, or solving a stochastic master equation \cite{flurin_using_2020}. In regards to shadow tomography methods, the comparison is less direct since the goal of these methods is not to reconstruct the dynamics at every time point, but rather to use a small set of measurements to predict a much larger set of specific observables.

A similar approach to ours for constructing a dynamical model of the noise channel using the Markovian Lindblad master equation was considered in~\cite{samach_lindblad_2021}. Here, in order to go beyond the Markovian approximation, a model based on a class of time-convolution-less master equations is considered, and a damping rate function describes the non-Markovian dynamics of the system at the sampling time point~\cite{xue_inverse-system_2020}. This method provides a discrete description of the noise dynamics but does not directly provide physical insight into quantifying the timescale of the memory effect. Likewise, a more sophisticated and systematic approach, the process tensor (PT) framework has been proposed~\cite{pollock_non-markovian_2018} and experimentally demonstrated~\cite{white_demonstration_2020,xiang_quantify_2021} to characterize non-Markovian dynamics on actual quantum processors.  The PT model, once fully characterized, can be used to interpolate the dynamics between discrete times due to the containment property of the PT map. However, its construction and the interpretation of the memory effect are fairly involved. 

Our method provides a continuous dynamical description of the channel beyond the Markov approximation, and does not require full process tomography to construct the PMME model. 
However, a disadvantage of our method is that it requires one to construct a parametric model; thus, it does not account for the most general noise process. Moreover, because of the specific parametrization that the PMME demands, the resulting optimization problem is no longer convex, i.e., a unique global minimum of the optimization problem is not guaranteed. Additionally, our method requires the consideration of a hierarchy of kernels to find a model that is complex enough to describe the data accurately. But since the PMME is straightforward to solve, model estimation under different kernels is straightforward so long as there is a model selection metric under which the best model among those candidates can be selected.  A promising future approach for retaining the physical interpretability and analytical solvability that come with using the PMME is the use of machine learning, especially in light of the recent development of neural ODE solvers~\cite{chen2019neural}, in order to avoid overfitting and structural errors of the PMME model.

Our work is a proof-of-principle demonstration of the PMME tomography protocol on a single qubit. An extension to the multi-qubit case is the natural next step. This will require a  convergence analysis to decide how the fitting data set scales as the number of qubits increases.  In this work, we have focused on the free evolution channel of a single IBMQE qubit, which we have shown to be highly susceptible to non-Markovian noise. As a next step, it would be interesting to extend the protocol to characterize non-Markovian effects during computation, in order  to understand whether such effects are significant beyond qubit idle times. By explicitly including non-Markovianity in the dynamical characterization and modeling, we expect that it will become possible to use these realistic noise models to improve and tailor error suppression and correction techniques, and ultimately realize high fidelity quantum control and computation.

\begin{acknowledgments}
This material is based upon work supported by the National Science Foundation the Quantum Leap Big Idea under Grant No.~OMA-1936388. HZ and BP acknowledge Namit Anand, Humberto Munoz Bauza, Jenia Mozgunov, Vinay Tripathi for insightful discussions. DAL also acknowledges support under a DOE/HEP QuantISED program grant, QCCFP/Quantum Machine Learning and Quantum Computation Frameworks (QCCFP-QMLQCF) for HEP, Grant No. DE-SC0019219.
We acknowledge the use of IBM Quantum services for this work. The views expressed are those of the authors, and do not reflect the official policy or position of IBM or the IBM Quantum team.
\end{acknowledgments}

\appendix
\setcounter{table}{0}
\renewcommand{\thefigure}{S\arabic{figure}}
%\section{Comparison with previous works}
%\label{app:A}
%
%
%A similar approach for constructing a dynamical model of the noise channel using the Markovian Lindblad master equation was considered in~\cite{samach_lindblad_2021}. Here, in order to go beyond the Markovian approximation, a model based on a class of time-convolution-less master equations is considered, and a damping rate function describes the non-Markovian dynamics of the system at the sampling time point~\cite{xue_inverse-system_2020}. This method provides a discrete description of the noise dynamics but does not directly provide physical insights into quantifying the timescale of the memory effect. Likewise, a more sophisticated and systematic approach, the process tensor (PT) framework has been proposed~\cite{pollock_non-markovian_2018} and experimentally demonstrated~\cite{white_demonstration_2020,xiang_quantify_2021} to characterize non-Markovian dynamics on real quantum processors.  The PT model, once fully characterized, can be used to interpolate the dynamics between discrete times due to the containment property of the PT map. However, its construction and the interpretation of the memory effect are fairly involved. 
%Our method provides a continuous dynamical description of the channel beyond the Markov approximation compared with the approaches above. Moreover, our method does not require full process tomography to construct the PMME model. 

\section{System information}
\label{app:sysinfi}

The IBMQE device used in this work is Athens, which is a 5-qubit processor consisting of superconducting transmon qubits. The main qubit we perform PMME tomography on is qubit 0 (Q0). The relevant device calibration details on the date of data collection are provided in Table~\ref{tab:backend}.
\begin{table}[h]
\begin{center}
\begin{tabular}{ |c|c|c|c|}
\hline
 Data set & Fig. \ref{fig:fit_and_test} & Fig. \ref{fig:app_zplusstate}, Fig. \ref{fig:app_zminusstate} & Fig. \ref{fig:app_xplusstate}  \\ 
 \hline
 Date collected &  6/25/2021 & 6/30/2021 & 7/1/2021 \\
 \hline
 $T_1$ ($\mu$s) & 72.6 & 70.8 & 75.2  \\  
 \hline
 $T_2$ ($\mu$s) & 93.4 & 82.6 & 62.9 \\
 \hline
readout error [$10^{-2}$]& 1.9 & 0.99 & 1.00 \\
\hline
\end{tabular}
\end{center}
\caption{Qubit calibration information of the  \textit{ibmq\_athens} processor on the date of data collection.}
\label{tab:backend}
\end{table}%

\section{Measurement error mitigation}
\label{app:MEM}

Measurement error mitigation is performed by using information from calibration experiments to remove any systematic bias in the measurement results~\cite{Temme:2017vc}. The calibration experiments involve preparing of computational basis states $| j \rangle$, which are then used to learn the response matrix $M$. The entries \( m_{kj} = \text{probability} (\text{prepare } | j \rangle | \text{measure bitstring } k ) \) represent conditional probabilities. Any subsequent experiment gives us the  measured probability vector $\vec{p}(E)$ which is used to infer the true probability of vector \(\vec{t}(E) = f(\vec{p}(E), M)\). The most commonly used MEM method -- called response matrix inversion method -- defines $\vec{t} = M^{-1} \vec{p}$.  Crucially, $M^{-1}$ is not stochastic, so $\vec{t}(E)$ can have negative entries. Recently,  a Bayesian solution to the non-stochasticity problem was proposed~\cite{nachman2020unfolding}.  In this method, inspired from similar unfolding methods in high-energy physics, we start with a prior truth spectrum $\vec{t}^0$ and update it using Bayes' rule to get
$$ t_i^{n+1} = \sum\limits_{j} \frac{M_{ji} t^{n }_{i}}{\sum\limits_{k} M_{jk} t_{k}^{n }} \times p_j. $$
The prior $t_{i}^{n}$ is updated using the response matrix $M$ and gives the posterior $t_{i}^{n+1}$, and the process proceeds for $100$ iterations (in practice this was found to always be sufficient for convergence). After each tomography experiment, the probabilities of each measurement outcome are updated using measurement error mitigation.

\section{Damping basis construction}
\label{app:A}

We present a systematic construction of the damping basis, sketched originally in Ref.~\cite{briegel_quantum_1993}. For our purposes, the damping basis is simply the basis of left and right eigenoperators (or eigenmatrices) $\{L_i\}$ and $\{R_i\}$, respectively, of the superoperator (e.g., Lindbladian) $\mathcal{L}$.

Let $\mc{H}$ denote a $d$-dimensional Hilbert space, $\mc{B}(\mc{H})$ the space of linear operators acting on $\mc{H}$, and consider a superoperator $\mc{L}:\mc{B}(\mc{H})\mapsto \mc{B}(\mc{H})$. We are interested in particular in superoperators in Lindblad form, i.e., Eq.~\eqref{eq:lindblad}:
\begin{align}
	\mathcal{L}(A)= -i[H, A] + \sum_{k} \gamma_{k}\left(V_{k} A V_{k}^{\dagger}-\frac{1}{2}\left\{V_{k}^{\dagger} V_{k}, A\right\}\right)\ ,
	\label{eq:Lindbladian}
\end{align} 
where $A,V_k\in \mc{B}(\mc{H})$, $H=H^\dag$, and $\gamma_{k}>0$ $\forall k$.

Let $\{F_i\}_{i=0}^{d^2-1}$ denote an orthonormal, Hermitian operator basis for $\mc{B}(\mc{H})$:
\begin{align}
\Tr(F_i F_j) = \d_{ij}\ , \quad F_i = F_i^\dag\ , \quad  F_0=I/\sqrt{d}\ .
\end{align}
For example, when $\mc{H} = \mathbb{C}^2$ (a qubit), we can choose the normalized Pauli matrices as the operator basis, i.e., $\{F_i\} = \{I,\s^x,\s^y,\s^z\}/\sqrt{2}$.

For any $A\in \mc{B}(\mc{H})$ we can then expand $A = \sum_i a_i F_i$, and in particular $\mc{L}(F_i) = \sum_j \ell_{ji} F_j$ (note the transposed index order), where $\ell$ is the matrix representation of $\mc{L}$ in the given basis, with the matrix elements given by $\ell_{ij} = \Tr[F_i\mc{L}(F_j)]$. The basis $\{F_i\}$ ``coordinatizes'' both $\mc{L}$ and the operators in $\mc{B}(\mc{H})$. 

Assume that the superoperator $\mc{L}$ satisfies the Hermiticity-preservation condition
\begin{align}
\left[\mathcal{L}\left(A\right)\right]^{\dagger}=\mathcal{L}\left(A^{\dagger}\right)\ .
\label{eq:Herm-pres}
\end{align}
This is true, in particular, for the Lindbladian~\eqref{eq:Lindbladian}, as is easily checked.
Let us show that then the matrix $\ell$ representing $\mc{L}$  in the chosen basis is real:
\begin{align}
\label{eq:L-real}
\ell_{ij}^*=\Tr\left(\left[\mathcal{L}\left(F_{i}\right)\right]^{\dagger}F_{j}^{\dagger}\right)=\Tr\left(\mathcal{L}\left(F_{i}\right)F_{j}\right)=\ell_{ij}\ .
\end{align}
Thus, after coordinatization the superoperator $\mathcal{L}$ can be seen as
a $d^2 \times d^2$ dimensional matrix $\ell\in \mathbb{R}^{d^{4}}$.

Now assume that $A$ is a right eigenoperator of $\mc{L}$ with eigenvalue $\lambda$, i.e., $\mc{L}(A) = \lambda A$. Then:
\begin{align}
\mc{L}(A) = \sum_{ij} a_i \ell_{ji} F_j = \lambda A = \lambda  \sum_i a_i F_i \ .
\end{align}
Taking the trace of both sides after multiplying from the right by $F_k$ yields $\sum_i a_i \ell_{ki} = \lambda a_k$, i.e., 
\begin{align}
\ell \vec{a} = \lambda \vec{a}\ ,
\end{align}
where $\vec{a} = (a_0,\dots,a_{d^2-1})^t$ is a column vector (the superscript $t$ denotes the transpose). Thus, if $R_i = \sum_j (\vec{r_i})_j F_j$ is a right eigenoperator of $\mc{L}$ then its coordinates-vector $\vec{r_i}$ is a right eigenvector of $\ell$. Conversely, by solving the linear algebra problem of finding the set of right eigenvectors $\{\vec{r_i}\}$ of $\ell$, we can construct the right eigenoperators of $\mc{L}$ as 
\begin{align}
\label{eq:Rfromr}
R_i = \sum_j (\vec{r_i})_j F_j \ .
\end{align}

Now consider the set of left eigenvectors $\vec{l_i}^t$ of $\ell$: $\vec{l_i}^t \ell = \lambda_i \vec{l_i}^t$. These are also the right eigenvectors of $\ell^t$: $\ell^t \vec{l_i} = \lambda_i \vec{l_i}$. Note that the left and right eigenvalues of $\ell$ are identical since the determinant of a matrix equals the determinant of its transpose. 

We define $\mc{L}^\dag$ as usual via the inner product relation
\begin{align}
\langle \mc{L}^\dag(A),B\rangle = \langle A,\mc{L}(B)\rangle\ ,
\label{eq:dag-def}
\end{align} 
where we use the Hilbert-Schmidt inner product 
\begin{align}
\langle A,B\rangle \equiv \Tr (A^\dag B)\ . 
\end{align}
Specifically, for the Lindbladian in Eq.~\eqref{eq:Lindbladian}, this implies that
\begin{align}
\mc{L}^\dag (A) = i[H, A] + \sum_{k} \gamma_{k}\left(V^{\dagger}_{k} A V_{k}-\frac{1}{2}\left\{V_{k}^{\dagger} V_{k}, A\right\}\right)\ ,
	\label{eq:Lindbladian-dag}
\end{align}
as can easily be verified by direct substitution of this form of $\mc{L}^\dag (A)$ into Eq.~\eqref{eq:dag-def}.

Let us show that $\ell^t$ is the matrix representation of $\mc{L}^\dag$. To do so, consider the expansion $\mc{L}^\dag (F_i) = \sum_{j}\tilde{\ell}_{ji} F_j$; we will show that in fact $\tilde{\ell} = \ell^t$. Indeed, on the one hand we have from Eq.~\eqref{eq:Lindbladian}:
\begin{align}
\label{eq:Lindbladian-dag-proof}
(\ell^t)_{ji} &= \ell_{ij} = \Tr[F_i\mc{L}(F_j)] = -i\Tr(F_i[H, F_j]) \\
&+ \sum_{k} \gamma_{k}\left(\Tr[F_i V_{k} F_j V_{k}^{\dagger}]-\frac{1}{2}\Tr[F_i\{V_{k}^{\dagger} V_{k}, F_j\}]\right)\ , \notag
\end{align}
and on the other hand we have from Eq.~\eqref{eq:Lindbladian-dag}:
\begin{align}
\tilde{\ell}_{ji} &= \Tr[F_j\mc{L}^\dag(F_i)] = i\Tr(F_j[H, F_i]) \\
&+ \sum_{k} \gamma_{k}\left(\Tr[F_jV_{k}^{\dagger} F_i V_{k}]-\frac{1}{2}\Tr[F_j\{V_{k}^{\dagger} V_{k}, F_i\}]\right) \notag \ ,
\end{align}
which is easily checked to be equal to the expression for $(\ell^t)_{ji}$ in Eq.~\eqref{eq:Lindbladian-dag-proof} by cycling operators under the trace. Thus, 
\begin{align}
\label{eq:A13}
\mc{L}^\dag (F_i) = \sum_{j}{\ell}_{ij} F_j\ ,
\end{align}
and the same reasoning that we used above for the right eigenvectors and eigenoperators now leads to the conclusion that if $L_i = \sum_j (\vec{l_i})_j F_j$ is a right eigenoperator of $\mc{L}^\dag$, i.e., a left eigenoperator of $\mc{L}$, then its coordinates-vector $\vec{l_i}$ is a left eigenvector of $\ell$. Conversely, by solving the linear algebra problem of finding the set of left eigenvectors $\{\vec{l_i}\}$ of $\ell$, we can construct the left eigenoperators of $\mc{L}$ as 
\begin{align}
\label{eq:Lfroml}
L_i = \sum_j (\vec{l_i})_j F_j \ .
\end{align}
Finally, it is well known that each left eigenvector is orthogonal to all right eigenvectors except its corresponding one (the one is shares an eigenvalue with), and vice versa~\cite{Press:1992uc}. By choice of normalization, the inner products of corresponding left and right eigenvectors can always be made unity for any matrix with nondegenerate eigenvalues. Assume nondegeneracy and that we have normalized $\ell$'s inner products of corresponding left and right eigenvectors, i.e., $\vec{l_i}\cdot \vec{r_j} = \d_{ij}$. Let us show in which sense this property is inherited by the left and right eigenoperators of $\mc{L}$:
\begin{align}
\Tr(L_i R_j) &= \sum_{kl} (\vec{l_i})_k (\vec{r_j})_l \Tr(F_k F_l) = \sum_k (\vec{l_i})_k (\vec{r_j})_k \notag \\
& = \vec{l_i}\cdot \vec{r_j} = \d_{ij}\ .
\label{eq:LR-ortho}
\end{align}
Note that $\Tr(L_i R_j) \neq \langle L_i,R_j\rangle$ since we do not take the Hermitian conjugate of $L_i$ under the trace [this only becomes possible if $\ell$ is symmetric, since we would need its eigenvalues to be real in order for $\Tr(L_i R_j) = \Tr(L^\dag_i R_j)$ to hold].

\section{Analytical solution of the PMME}
\label{app:B}

Here we present the analytical solution of the PMME, Eq.~\eqref{eq:pmme}, for our model.
We take the Laplace transform, and the PMME becomes:
\begin{align}
	s\tilde{\rho}(s)-\rho(0)&=\mathcal{L}_0\tilde{\rho}(s) +  \\
	&\qquad \mathcal{L}_1 \operatorname{Lap}\left[k(t)\operatorname{exp}\left(\mathcal{L}_0+\mathcal{L}_1\right)t\right]\tilde{\rho}(s) \ . \notag
\end{align}
To deal with $e^{\mathcal{L}t}:=e^{\left(\mathcal{L}_0+\mathcal{L}_1\right)t}$, it is convenient to work in the damping basis of $\mathcal{L}$, as defined in Appendix~\ref{app:A}. Recall that the sets of right and left eigenoperators of $\mathcal{L}$, $\{R_i\}$ and $\{L_i\}$, are complete and mutually orthonormal in the sense of Eq.~\eqref{eq:LR-ortho}. 
We therefore expand $\rho$ in the basis of right eigenoperators of $\mathcal{L}$:
\begin{align}
\label{eq:rho}
	\rho(t)=\sum_{i} \mu_{i}(t) R_{i}\ ,
\end{align} 
where the expansion coefficients are 
\begin{align}
	\mu_{j}(t)=\sum_{i} \mu_{i}(t) \Tr\left(L_{j} R_{i}\right)=\Tr\left[L_{j} \rho(t)\right]\ .
\end{align}
Substituting Eq.~\eqref{eq:rho} into the PMME Eq.~\eqref{eq:pmme}, we obtain:
\begin{align}
\label{eq:B4}
	&\sum_i\frac{\partial\mu_i(t)}{\partial t}R_i=\sum_i \mu_i(t) \mathcal{L}_0 R_i +\\
	&\qquad \sum_i \int_0^t dt^\prime k(t^\prime)\operatorname{exp}\left(\lambda_it\right)\mu_i(t-t^\prime)\mathcal{L}_1R_i\ . \notag
\end{align}
Notice that if we assume that $\left[\mathcal{L}_0,\mathcal{L}_1\right]=0$ (as is the case for us), then $\mathcal{L}_0$ and $\mathcal{L}_1$ both commute with $\mathcal{L} = \mathcal{L}_0+\mathcal{L}_1$ and hence share the same set of left and right eigenoperators with it, i.e., $\mathcal{L}_0 (R_i)=\lambda_i^0 R_i$, $\mathcal{L}^\dag_0(L_i)=\lambda_i^0 L_i$, $\mathcal{L}_1 (R_i)=\lambda_i^1 R_i$, $\mathcal{L}^\dag_1(L_i)=\lambda_i^1 L_i$.
Multiplying both sides of Eq.~\eqref{eq:B4} by $L_j$ from left and taking the trace, we obtain, under this assumption:
\begin{align}
	\frac{\partial\mu_i(t)}{\partial t}=\lambda_i^0\mu_i(t) + \lambda_i^1\int_0^t dt^\prime k(t^\prime)\operatorname{exp}\left[\lambda_it^\prime\right]\mu_i(t-t^\prime)\ .
\end{align}
Take the Laplace transform of both sides and use the shifting property of the Laplace transform, we have:
\bes
\begin{align}
	s\tilde{\mu}(s)-\mu_i(0) &= \lambda_i^0 \tilde{\mu}_i(s) + \lambda_i^1 \operatorname{Lap}\left[k(t)e^{\lambda_it}\right]\tilde{\mu}_i(s)
	\\&= \lambda_i^0 \tilde{\mu}_i(s) + \lambda_i^1 \tilde{k}(s-\lambda_i)\tilde{\mu}_i(s).
\end{align}
\ees
Therefore,
\begin{align}
	\tilde{\mu}_i(s)=\frac{1}{s-\lambda_i^0-\lambda_i^1k(s-\lambda_i)}.
\end{align}
Taking the inverse Laplace transform:
\begin{align}
	\mu_i(t)=\xi_i(t) \mu_i(0),	
\end{align}
where:
\bes
\begin{align}\label{eq:xi}
\xi_i(t) &=\operatorname{Lap}^{-1}\left[\frac{1}{s-\lambda_i^0-\lambda_i^1k(s-\lambda_i)}\right] \\
\mu_i(0) &=\Tr\left[L_i\rho(0)\right].	
\end{align}
\ees

\subsection{Solution with the specific Lindbladian and kernels}

Choosing the operator basis as $\{F_i\} = \{I,\s^x,\s^y,\s^z\}/\sqrt{2}$, we follow the methodology of Appendix~\ref{app:A} and find the matrix representation of $\mc{L} = \mc{L}_0+\mc{L}_1$ of the specific PMME model we seek to construct in Sec.~\ref{sec:model_motivations} to be
\begin{align}
\ell = 
\left(
\begin{array}{cccc}
 0 & 0 & 0 & 0 \\
 0 & -\frac{\Gamma _s}{2}-2 \gamma _z & -\omega_z & 0 \\
 0 & \omega_z & -\frac{\Gamma_s}{2}-2 \gamma_z & 0 \\
 \gamma _- - \gamma _+ & 0 & 0 & -\Gamma _s \\
\end{array}
\right)\ .
\end{align}
The eigenvalues of $\ell$ are:
\begin{align}
\{\lambda_i\}&=\{0,-\frac{1}{2}\Gamma_s-2\gamma_z+i\omega_z,-\frac{1}{2}\Gamma_s-2\gamma_z-i\omega_z,-\Gamma_s\}\ .
\end{align}
It is straightforward to compute and normalize the corresponding right and left eigenvectors $\{\vec{r_i},\vec{l_i}\}_{i=1}^4$ such that they are mutually orthonormal, i.e., $\vec{l_i}\cdot \vec{r_j} = \d_{ij}$. This allows us to find the right and left eigenoperators of $\mc{L}$ using Eqs.~\eqref{eq:Rfromr} and~\eqref{eq:Lfroml}, which yields:
\bes
\label{eq:damping_basis}
\begin{align}
R_1&=\begin{pmatrix}
\frac{1}{\Gamma_r+1} & 0 \\
0 & \frac{\Gamma_r}{\Gamma_r+1}
\end{pmatrix},\
R_2=\begin{pmatrix}
0 & 1 \\
0 & 0	
\end{pmatrix},\ 
R_3=\begin{pmatrix}
0 & 0 \\
1 & 0	
\end{pmatrix},\\
R_4&=\begin{pmatrix}
	-\frac{1}{\Gamma_r+1} & 0 \\
	0 & \frac{1}{\Gamma_r+1}
\end{pmatrix}\,\\
L_1&=\begin{pmatrix}
1 & 0 \\
0 & 1	
\end{pmatrix},\ 
L_2=\begin{pmatrix}
0 & 0 \\
1 & 0	
\end{pmatrix}, \
L_3=\begin{pmatrix}
	0 & 1 \\
	0 & 0
\end{pmatrix},\\
L_4&=\begin{pmatrix}
-\Gamma_r & 0 \\
0 & 1
\end{pmatrix}\ .
\end{align}
\ees
It is simple to verify that this set satisfies Eq.~\eqref{eq:LR-ortho} as required.

Applying $\mathcal{L}_0$ and $\mathcal{L}_1$ to $\{R_i\}$, we find the corresponding sets of eigenvalues:
\bes
\begin{align}
\{\lambda_i^0\}&=\{0,-\frac{\Gamma_s}{2}+i\omega_z,-\frac{\Gamma_s}{2}-i\omega_z,-\Gamma_s\}\\
\{\lambda_i^1\}&=\{0,-2\gamma_z,-2\gamma_z,0\} 
\end{align}
\ees
Next, we need to evaluate Eq.~\eqref{eq:xi} with the specific forms of the kernels we have chosen.
Regardless of the kernel, 
\bes
\begin{align}
\tilde{\xi}_1(s)=\frac{1}{s} & \Longleftrightarrow \xi_1(t)=1 \\
\tilde{\xi}_4(s)=\frac{1}{s+\Gamma_s}  & \Longleftrightarrow  \xi_4(t) = e^{-\Gamma_st}\ ,
\end{align}
\ees
while for $i=2,3$:
\bes
\begin{align}
f(t) &\equiv \xi_2(t) =\xi_3^*(t)\\
&=\operatorname{Lap}^{-1}\left[\frac{1}{s-\lambda_2^0-\lambda_2^1\tilde{k}(s-\lambda_2)}\right]\ .
\end{align}
\ees

For the exponentially decaying kernel in Eq.~\eqref{eq:kernel1}, we make the parameter substitution $x\equiv 2+\frac{b_0}{\gamma_z}$ and $y\equiv \frac{\gamma_s}{2\gamma_z}-i\frac{\omega_z}{\gamma_z}$ and transform the variables in the Laplace transform correspondingly as $s=\gamma_zz$ and $\tau=\gamma_zt$, to get:
\bes
\begin{align}
	f(\tau)&=\operatorname{Lap}^{-1}\left[\tilde{f}(z)\right]\\
	\tilde{f}(z)&=\frac{z+x+y}{(z+y)^2+x(z+y)+2}\ .
\end{align}
\ees
The analytical solution can be found by using the residue theorem~\cite{complex:book}:
\begin{align}
	f(\tau)=\text{sum of residues of } e^{zt}\tilde{f}(z)\text{ at poles of } \tilde{f}(z)\ .
	\label{eq:RT}
\end{align}
The rational function $\tilde{f}(z)$ has two poles $z_1$ and $z_2$
\begin{align}
	z_{1,2}=\frac{1}{2}\left(-x-2y\pm\sqrt{D}\right)\ , \quad D\equiv x^2-8\ ,
\end{align}
and using Eq.~\eqref{eq:RT}:
\begin{align}
	f(\tau)=e^{z_1\tau}\frac{z_1+y+x}{z_1-z_2}+e^{z_2\tau}\frac{z_2+y+x}{z_2-z_1} \ .
\end{align}

For the other kernel in Eq.~\eqref{eq:kernel2}, we make the parameter substitution $a_0\equiv \gamma_z^2 x$, $y\equiv\frac{\gamma_s}{2\gamma_z}-i\frac{\omega_z}{\gamma_z}+2$, $a_1\equiv=\gamma_zw$,$b_0\equiv \gamma_z^2u$, $b_1\equiv \gamma_zv$, and transform the variables in the Laplace transform correspondingly as $s=\gamma_zz$ and $\tau=\gamma_zt$, to get:
\bes
\begin{align}
	f(\tau)&=\operatorname{Lap}^{-1}\left[\tilde{f}(z)\right]	\\
	\tilde{f}(z)&=\frac{p_1(z)}{\left(z+y-2\right)p_1(z)+2w\left(z+y\right)+2x} \ ,
\end{align}
\ees
where $p_1(z)=\left(z+y\right)^2+v(z+y)+u$. The analytical solution can therefore be found in terms of the roots of the cubic polynomial
\bes
\begin{align}
&\left(z+y-2\right)p_1(z)+2w\left(z+y\right)+2x\\
&\quad =z^3+c_2z^2+c_1z+c_0=0\ ,
\end{align}
\ees
where the coefficients are
\bes
\begin{align}
	c_2 &= 3y+v-2\\
	c_1 &= 3y^2+2vy-4y+3w-2v \\
	c_0 & = y^3+vy^2-2y^2+2wy-2vy+2x-2w\ .
\end{align}
\ees
The corresponding depressed cubic is found by the substitution $z=z^\prime-c_2/3$,
\begin{align}
r\left(z^\prime\right) = z^{\prime 3}+pz^\prime-q=0
\end{align}
where the coefficients are
\begin{align}
	p &= \frac{3c_1-c_2^2}{3}\\
	q &= \frac{9c_1c_2-27c_0-2c_2^3}{27}
\end{align}
which yields the cubic discriminant
\begin{align}
D=\left(\frac{p}{3}\right)^3+\left(\frac{q}{2}\right)^2
\end{align}
Further denote
\begin{align}
	S=\sqrt[3]{\frac{q}{2}+\sqrt{D}}, \quad T=\sqrt[3]{\frac{q}{2}-\sqrt{D}},
\end{align}
The zeros of the cubic are
\begin{align}
z_{1} &= -\frac{2 x}{3}+(S+T) \\
z_{2} &= -\frac{2 x}{3}-\frac{1}{2}(S+T)+\frac{i}{2} \sqrt{3}(S-T) \\
z_{3} &= -\frac{2 x}{3}-\frac{1}{2}(S+T)-\frac{i}{2} \sqrt{3}(S-T)
\end{align}
This completes the exact solution of the PMME.  

\begin{figure*}[t]
  \centering
  \includegraphics[width=\linewidth]{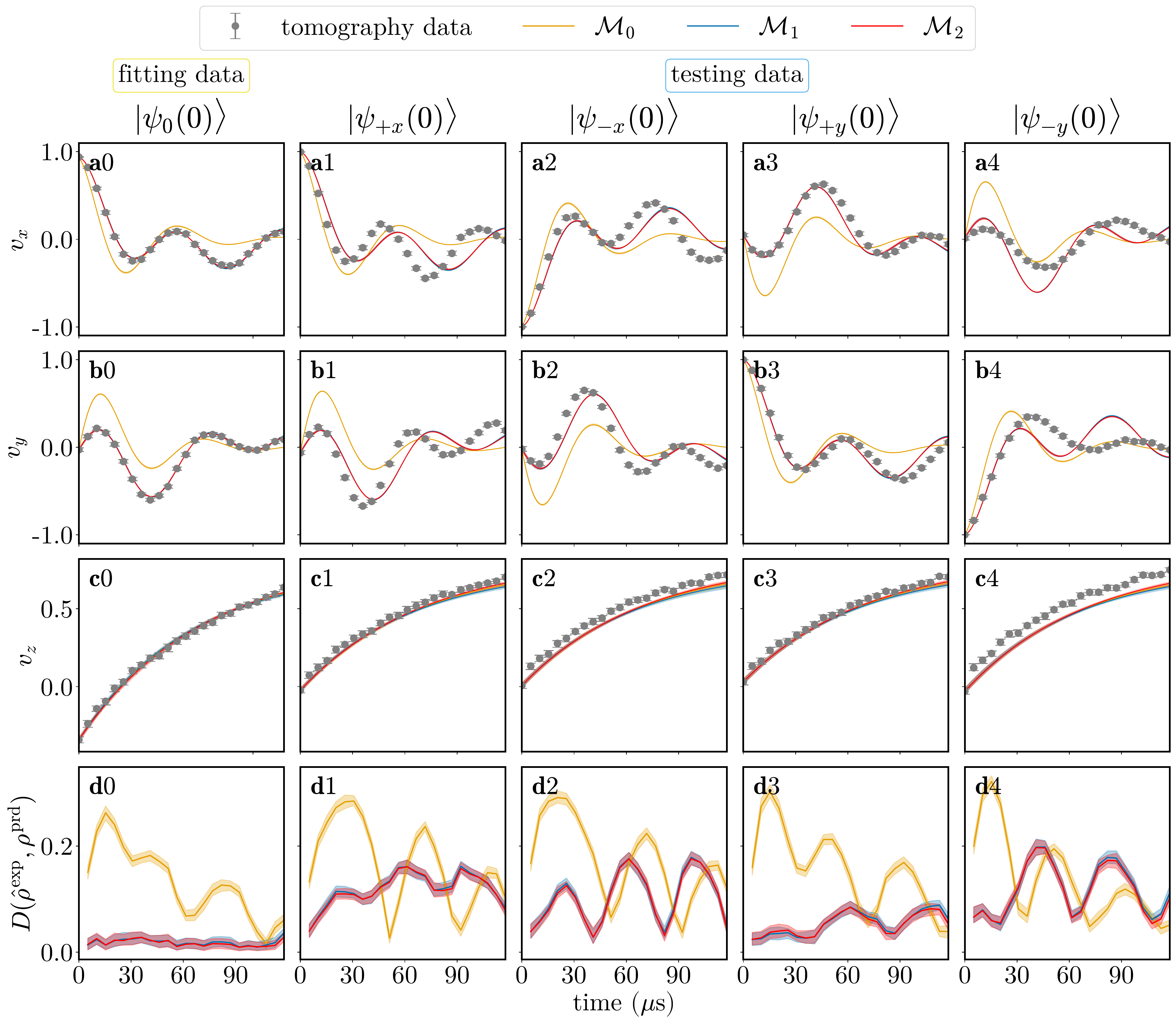}
  \caption{The tomography data set and the corresponding model predictions used to calculate the degree of non-Markovianity in Fig.~\ref{fig:nonMark} (a,b) when the spectator qubits are initialized in the ground state. (a0-c0) The fitting data set with the qubit initialized in $|\psi_0(0)\rangle$ that is used to find the best-fit Lindblad model $\mathcal{M}_0$ (orange lines), the PMME model with the type 1 kernel $\mathcal{M}_1$, and the PMME model with the type 2 kernel $\mathcal{M}_2$. (a1-c2) The tomography data set with qubit initialized in $\ket{\psi_{+x}}$, $\ket{\psi_{-x}}$ and the prediction from the best-fit models from the fitting data set in (a0-c0). The data sets are used to evaluate the degree of non-Markovianity in Fig.~\ref{fig:nonMark} (a,b). (a3-c4) The tomography data set with the qubit initialized in $\ket{\psi_{+y}}$, $\ket{\psi_{-y}}$, and the prediction from the best-fit models from the fitting data set in (a0-c0). The data sets are used to evaluate the degree of non-Markovianity in Fig.~\ref{fig:nonMarkY} (a,b). (d0-d4) The distance between the tomographically constructed state and the state predicted by the best-fit models.}
  \label{fig:app_zplusstate}
\end{figure*}

\begin{figure*}[t]
  \centering
  \includegraphics[width=\linewidth]{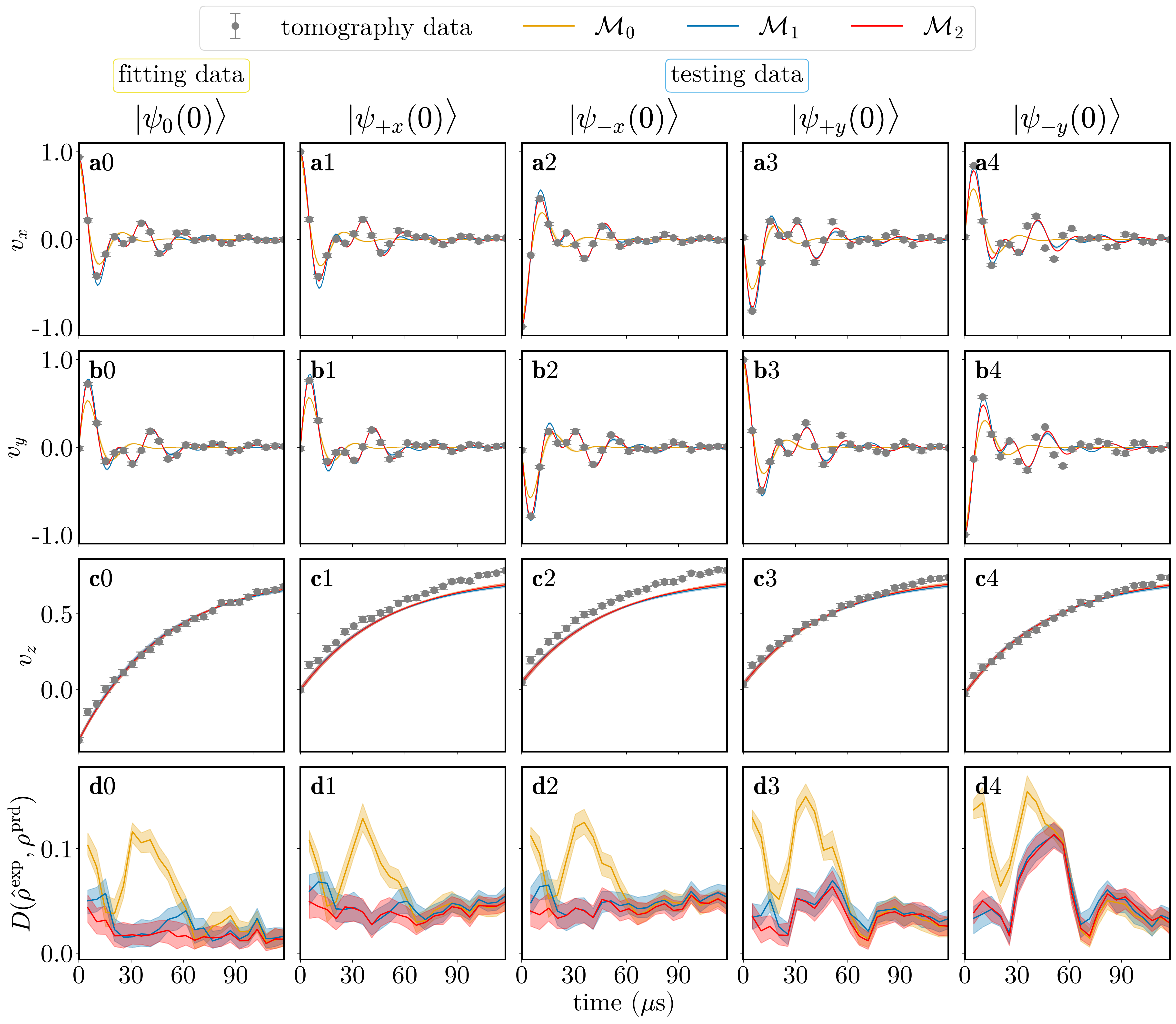}
  \caption{The tomography data set and the corresponding model predictions used to calculate the degree of non-Markovianity in Fig.~\ref{fig:nonMark} (c,d) when the spectator qubits are initialized in the excited state. (a0-c0) The fitting data set with the qubit initialized in $|\psi_0(0)\rangle$ that is used to find the best-fit Lindblad model $\mathcal{M}_0$ (orange lines), the PMME model with the type 1 kernel $\mathcal{M}_1$, and the PMME model with the type 2 kernel $\mathcal{M}_2$. (a1-c2) The tomography data set with qubit initialized in $\ket{\psi_{+x}}$, $\ket{\psi_{-x}}$ and the prediction from the best-fit models from the fitting data set in (a0-c0). The data sets are used to evaluate the degree of non-Markovianity in Fig.~\ref{fig:nonMark} (c,d). (a3-c4) The tomography data set with the qubit initialized in $\ket{\psi_{+y}}$, $\ket{\psi_{-y}}$, and the prediction from the best-fit models from the fitting data set in (a0-c0). The data sets are used to evaluate the degree of non-Markovianity in Fig.~\ref{fig:nonMarkY} (c,d). (d0-d4) The distance between the tomographically constructed state and the state predicted by the best-fit models.}
    \label{fig:app_zminusstate}
\end{figure*}

\section{Complete positivity of the PMME}
\label{app:cptp}

The complete positivity of the PMME is not guaranteed because of the freedom in choosing the kernel $k(t)$. A complete positivity test for the PMME was provided in Ref.~\cite{shabani_completely_2005}. Below we apply this test to the kernels specified in the model $\mathcal{M}_i$.

The solution of the PMME can be viewed as a map $\Phi$ acting on the operators represented by $d\times d$ matrices, where $d$ is the dimension of the Hilbert space $\mathcal{H}=\text{span}\{|i\rangle\}_{i=1}^d$. Using Eq.~\eqref{eq:rho},
\bes
\begin{align}
\rho(t)&=\sum_{i} \mu_{i}(t) R_{i}=\sum_{i} \xi_{i}(t) \mu_{i}(0) R_{i}\\
&=\sum_{i} \xi_{i}(t) \Tr\left[L_{i} \rho(0)\right] R_{i}=\Phi[\rho(0)],
\end{align}
\ees
where 
\begin{align}
\Phi[X] \equiv \sum_{i} \xi_{i}(t) \Tr\left[L_{i} X\right] R_{i}.
\end{align}
Let $|\phi\rangle=\sum_i |i\rangle\otimes | i\rangle $ be a maximally entangled state in $\mathcal{H}\otimes\mathcal{H}$. According to the Choi's theorem~\cite{choi_completely_1975}, $\Phi$ is CP if and only if the Choi matrix $C \geq 0$, where 
\begin{align}
C=(\mathbb{I}\otimes\Phi)|\phi\rangle\langle \phi|=\sum_{i j}|i\rangle\langle j| \otimes \Phi[|i\rangle\langle j|]\ .
\end{align}
We construct the Choi matrix for the PMME. Let us pick the basis state $|i\rangle$ to be a column vector of zeros, except for a $1$ in position $i$; we have:
\bes
\begin{align}
C&=\sum_{i j}|i\rangle\langle j| \otimes \sum_{k} \xi_{k}(t) \Tr\left[L_{k}|i\rangle\langle j|\right] R_{k} \\
&= \sum_{k} \xi_{k}(t) \sum_{i j}|i\rangle\langle j| \otimes\left\langle j\left|L_{k}\right| i\right\rangle R_{k} \\
&= \sum_{k} \xi_{k}(t) \sum_{i j}|i\rangle\langle j|\left(L_{k}^{T}\right)_{i j} \otimes R_{k}
\end{align}
\ees
Hence,
\begin{align}
C=\sum_{k} \xi_{k}(t) L_{k}^{T} \otimes R_{k}>0
\end{align}
is the complete positivity condition for the kernel, under a given Lindbladian $\mathcal{L}$ and its set of left and right eigenvectors.

For the Lindbladian $\mathcal{L}$ in Eq.~\eqref{eq:lindblad} and its set of left and right eigenvectors in Eq.~\eqref{eq:damping_basis}, the Choi matrix is:
\begin{align}
C=\begin{pmatrix}
\frac{1+\Gamma_r \xi_4}{1+\Gamma_r} & 0 & 0 & \xi_2 \\
0&                             \frac{\Gamma_r(1-\xi_4)}{1+\Gamma_r} & 0 & 0 \\
0 & 0 & \frac{1-\xi_4}{1+\Gamma_r} & 0 \\
\xi_3 & 0 & 0 & \frac{\Gamma_r + \xi_4}{1+\Gamma_r}
\end{pmatrix}.
\end{align}
Its eigenvalues are found to be:
\bes
\begin{align}
	\lambda_1^c&=\frac{1-\xi_4}{1+\Gamma_r},\\
	\lambda_2^c &= \frac{\Gamma_r(1-\xi_4)}{1+\Gamma_r}\\
	\lambda_{3,4}^c &= \frac{1+\xi_4}{2}\pm \\
	& \sqrt{\left(\frac{\xi_4+1}{2}\right)^2-\frac{\Gamma_r+\xi_4+\Gamma_r^2\xi_4+\Gamma_r\xi_4^2-|\xi_2|^2}{(\Gamma_r+1)^2}}.
\end{align} 
\ees
Therefore, the PMME in this case corresponds to a CP map iff:
\bes
\begin{align}
	|\xi_4| &< 1,\\
	|\xi_2|=|\xi_3|&<\frac{\sqrt{(\Gamma_r+\xi_4)(1+\Gamma_r \xi_4)}}{1+\Gamma_r} < 1\ ,
\end{align}
\ees
which is a condition on the problem parameters $\gamma_z$, $\gamma_+$, $\gamma_-$ and the kernel parameters $\vec{a}$ and $\vec{b}$.

\begin{figure*}[t]
  \centering
  \includegraphics[width=\linewidth]{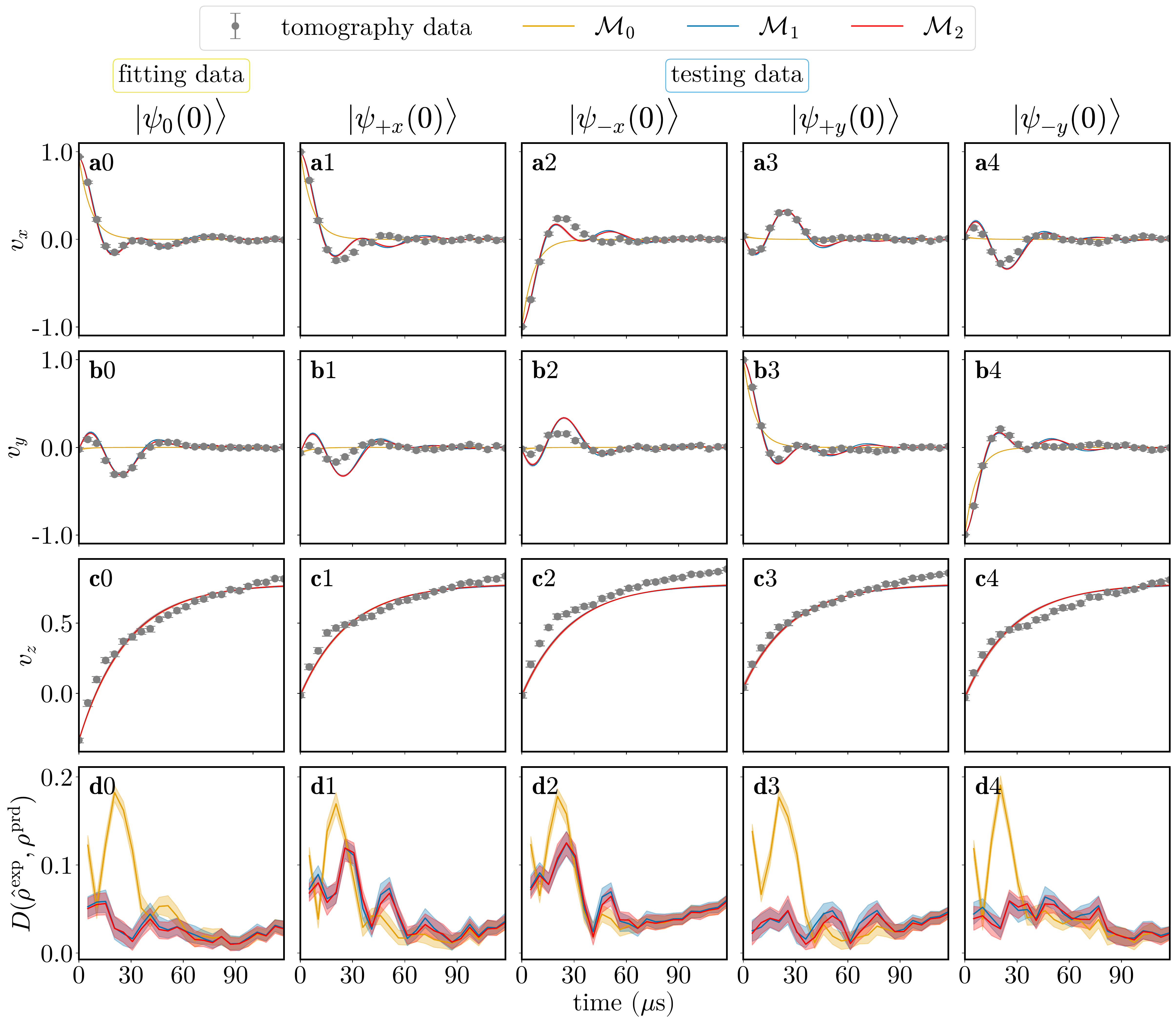}
  \caption{The tomography data set and the corresponding model predictions used to calculate the degree of non-Markovianity in Fig.~\ref{fig:nonMark} (e,f) when the spectator qubits are initialized in the excited state. (a0-c0) The fitting data set with the qubit initialized in $|\psi_0(0)\rangle$ that is used to find the best-fit Lindblad model $\mathcal{M}_0$ (orange lines), the PMME model with the type 1 kernel $\mathcal{M}_1$, and the PMME model with the type 2 kernel $\mathcal{M}_2$. (a1-c2) The tomography data set with qubit initialized in $\ket{\psi_{+x}}$, $\ket{\psi_{-x}}$ and the prediction from the best-fit models from the fitting data set in (a0-c0). The data sets are used to evaluate the degree of non-Markovianity in Fig.~\ref{fig:nonMark} (e,f). (a3-c4) The tomography data set with the qubit initialized in $\ket{\psi_{+y}}$, $\ket{\psi_{-y}}$, and the prediction from the best-fit models from the fitting data set in (a0-c0). The data sets are used to evaluate the degree of non-Markovianity in Fig.~\ref{fig:nonMarkY} (e,f). (d0-d4) The distance between the tomographically constructed state and the state predicted by the best-fit models.}
    \label{fig:app_xplusstate}
\end{figure*}

\section{PMME model construction result for different spectator qubit states}
\label{app:D}
\begin{figure}[h!]
  \centering
  \includegraphics[width=\linewidth]{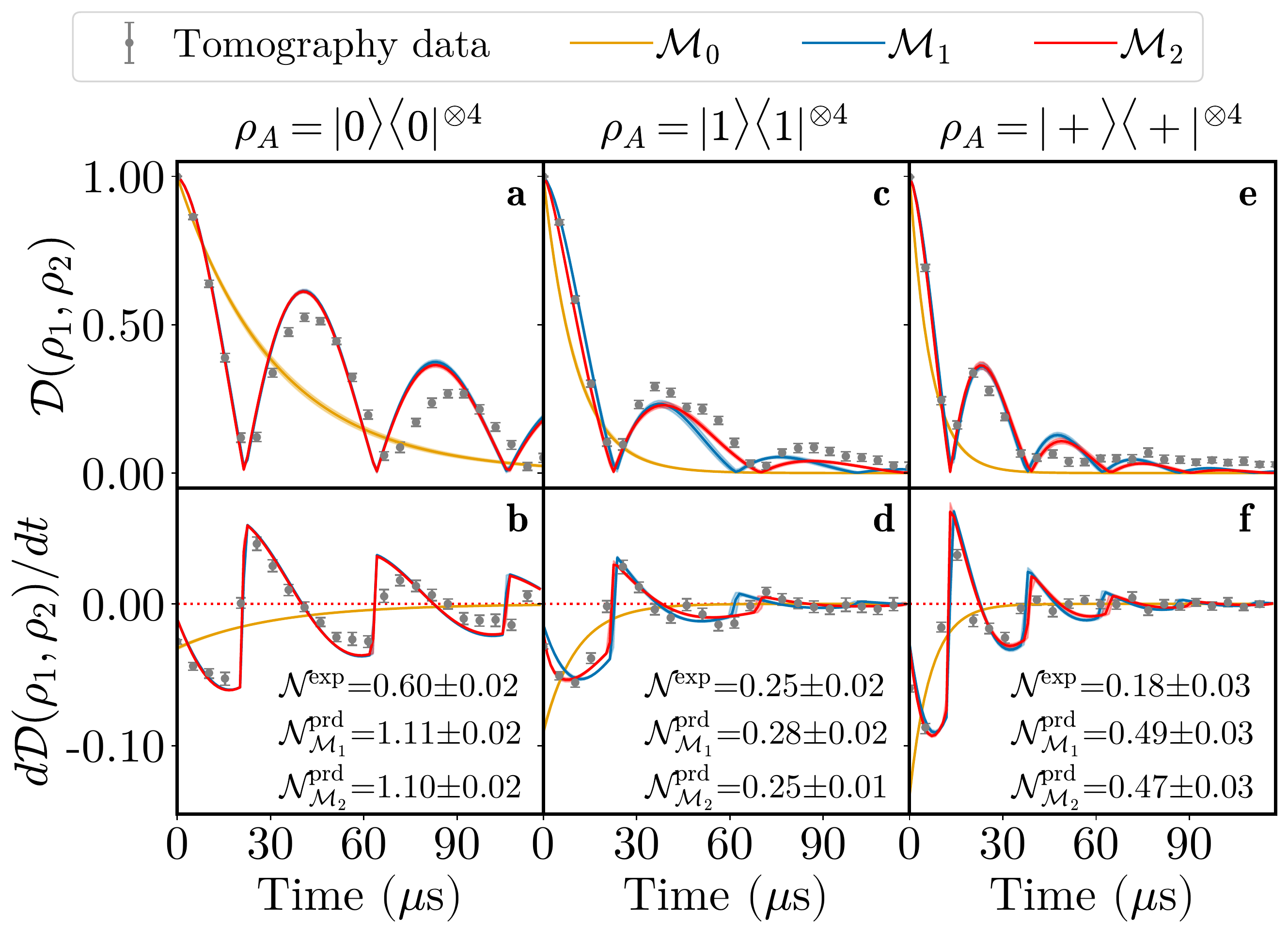}
  \caption{Non-Markovianity of qubit free-evolution dynamics for spectator qubits in the ground state (a,b), the excited state (c,d) and the $|+\rangle$ state (e,f). (a,c,e) The trace-norm distance $\mathcal{D}(\rho_1(t),\rho_2(t))$ predicted by the best-fit models (solid lines) and experimentally measured by performing free-evolution tomography with a pair of initial states $\rho_1(0)=\ketb{+i}{+i}$ and $\rho_2(0)=\ketb{-i}{-i}$ (grey circles). (b,d,f) The derivative $\sigma(t)$, defined in Eq.~\eqref{eq:derivative}, predicted by the best-fit models (solid lines), and approximated experimentally  using forward differencing based on the tomography data in (a,c,d) (grey circles).}
    \label{fig:nonMarkY}
\end{figure}

In this section we present data supplementing the results of the PMME model construction reported in Fig.~\ref{fig:fit_and_test}, for the following three initial states of the spectator qubits: ground state $\ket{0}$, the excited state $\ket{1}$, and the $|+\rangle$ state. This data was also used to calculate the degree of non-Markovianity reported in Fig.~\ref{fig:nonMark}, which used different initial states of the main qubit. Figures~\ref{fig:app_zplusstate},~\ref{fig:app_zminusstate} and~\ref{fig:app_xplusstate} show the fitting data used to construct the model and the testing data to validate it. The initial states in the testing data sets are $\{\ket{\psi_{+x}}=|+\rangle, \ket{\psi_{-x}}=|-\rangle, |\psi_{+y}\rangle=|+i\rangle, |\psi_{-y}\rangle=|-i\rangle\}$ and they are used to evaluate the degree of non-Markovianity as in Eq.~\eqref{eq:derivative}, plotted in Figs.~\ref{fig:nonMark} and~\ref{fig:nonMarkY}. The initial state pairs $\rho_1(0)=\ketb{+}{+}$ and $\rho_2(0)=\ketb{-}{-}$ (or $\rho_1(0)=\ketb{+i}{+i}$ and $\rho_2(0)=\ketb{-i}{-i}$) are optimal pairs such that they feature a maximal flow of information from the environment back to the system \cite{wissmann_optimal_2012}.

\begin{figure}[t]
  \centering
  \includegraphics[width=\linewidth]{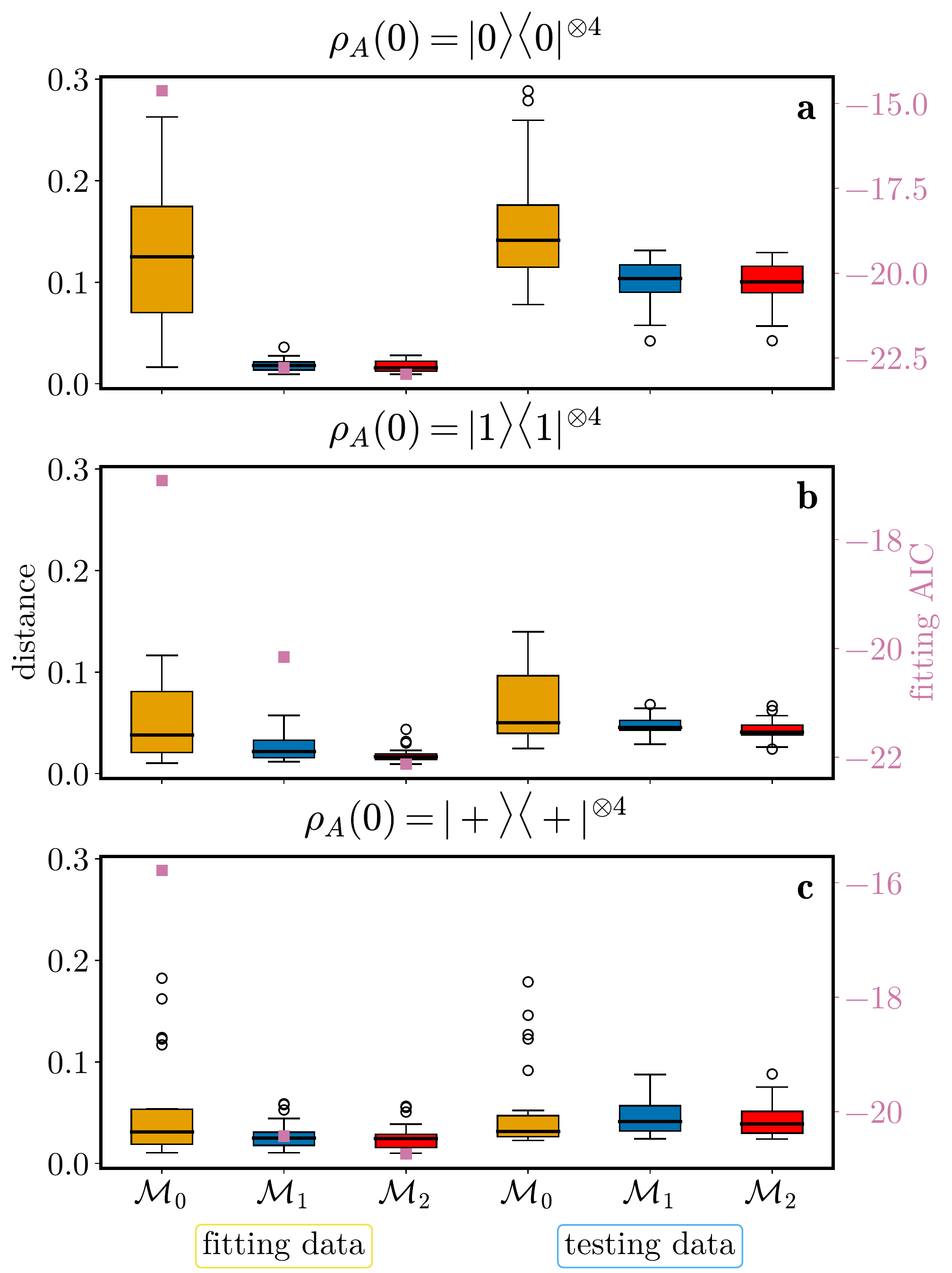}
  \caption{Summary of the results of running the PMME tomography protocol on the IBMQE processor ibmq\_athens, showing how well the model candidates describe the fitting data set and the testing data set in Fig.~\ref{fig:app_zplusstate} for the spectator qubits in the ground state (a), for those in Fig.~\ref{fig:app_zminusstate} for the spectator qubits in the excited state (b) and for those in Fig.~\ref{fig:app_xplusstate} for the spectator qubits in the $|+\rangle$ state (c). The box plots show the trace norm distance (see text) and the median is reported as the middle value of $\{\mathcal{D}_j\}$. The lower line of the box corresponds to the lower quartile of the data (25th percentile, Q1), and the upper line of the box corresponds to the upper quartile of the data (75th percentile, Q3). Let IQR denote the interquartile range: IQR = Q3-Q1. The outliers, plotted in circles, are the data outside the range (Q1-1.5*IQR, Q3+1.5*IQR). 
}
    \label{fig:app_box}
\end{figure}

The yellow, blue, and red solid lines in Fig.~\ref{fig:app_zplusstate}, Fig.~\ref{fig:app_zminusstate} and Fig.~\ref{fig:app_xplusstate} represent the constructed models $\mathcal{M}_0$, $\mathcal{M}_1$ and $\mathcal{M}_2$, respectively. On the fitting data set, we find that the Lindblad model $\mathcal{M}_0$ does not adequately describe the data, while the PMME models $\mathcal{M}_1$ and $\mathcal{M}_2$ describe the data accurately. On the testing data set, the constructed models provide a qualitatively adequate prediction for different initial states, but deviations do arise relative to the the experimentally constructed states (see Fig.~\ref{fig:app_zplusstate}-\ref{fig:app_xplusstate} columns 1, 2, and 4). Due to system fluctuations, which the PMME models cannot capture, the model constructed from the fitting data set may lose some of its predictive power for the testing data set. Examples of such system fluctuations include fluctuations of qubit $T_1$ relaxation time and 1/$f$ noise in the qubit frequency.

Nonetheless, compared with the Lindblad model, the PMME models provide higher levels of agreement for the fitting data sets and a more accurate prediction for the testing data sets in all spectator qubit configurations.  This can be seen in the bottom row of Figs.~\ref{fig:app_zplusstate}-\ref{fig:app_xplusstate}, and is summarized in Fig.~\ref{fig:app_box} which compares the models using the Akaike Information Criterion (AIC) and the trace distance metric discussed below.  

To compare the goodness of the fit of the models and  the simplest model that best describes the fitting data set, we use the Akaike information Criterion (AIC) \cite{akaike_new_1974} which is defined as:
\begin{align}
	\text{AIC}=-2 \operatorname{ln} \left(\hat{L}(\hat{\theta} \vert D)\right)+2 p\ ,
	\label{eq:AIC}
\end{align}
where $\hat{L}$ denotes the likelihood function and $p$ is the number of free model parameters. The second term in the AIC ($2p$) is called the bias term and it penalizes the models with higher complexity. 
The AIC can be interpreted as the ``information loss'' (in the Kullback-Leibler divergence sense) of using some candidate model to approximate the ``true'' model. Akaike showed that the maximized log-likelihood in the first term of Eq.~\eqref{eq:AIC} is a biased estimator, and that under certain assumptions, the bias correction approximately equals $p$. The calculation of the first term in AIC depends strongly on the sample size used, and the bias-correction term becomes exact when the sample size diverges. For these reasons, the numerical value of AIC has no intrinsic significance as such.
With these caveats in mind, we can use the AIC metric to find a model that accurately describes the data and at the same time avoids overfitting, as lower (more negative) values indicate a better model.

The likelihood function $\hat{L}$ is defined over the observed dataset, and quantifies the likelihood of observing the data set as a function of the model parameters $\vec{\theta}$. It measures how well the data supports that particular choice of parameters. Since each tomography sample $\hat{\rho}$ is independently drawn, the likelihood $\hat{L}$ is a product of conditional probabilities:
\begin{align}
	\hat{L}(\vec{\theta}|\mathcal{D})=
	\prod_i\prod_k p_k(t_i;\vec{\theta})\ ,
\end{align}
where $p_k$ is the probability of observing the data point in measurement basis $k$ at time $t_i$.

We report the AIC values computed in this manner for the fitting data using the three models in Fig.~\ref{fig:app_box} (pink squares), for three different initial states of the ancilla qubits. However, as mentioned above, the numerical value of the AIC is not intrinsically meaningful. In practice, it is convenient to scale AICs with respect to the minimum AIC value among all models:
\begin{align}
	\Delta_{i}=\mathrm{AIC}_{i}-\min _{i} \mathrm{AIC}_{i}	
\end{align}
where $\min _{i} \mathrm{AIC}_{i}$ is the AIC value of the best model in the set. As see in Fig.~\ref{fig:app_box}, we find that the minimum is achieved for the $\mc{M}_2$ model. The AIC difference $\Delta_i$ estimates the information loss when using model $i$ rather than the estimated best model. Hence, the larger $\Delta_i$, the less plausible is model $i$. Some guidelines for the interpretation of AIC difference in the case of nested models are given in \cite{burnham_model_2002}, as summarized in Table~\ref{tab:3}. Given the much larger $\Delta_0=\mathrm{AIC}_0-\mathrm{AIC}_1$ value, we find that the data is considerably less in favor of the Lindblad model $\mathcal{M}_0$, despite the fact that it has the smallest number of free parameters.

Another metric we use here for comparison is the trace-norm distance between the experimentally constructed state and the model predicted state during its evolution. Let $D_j^k$ denote the distance between the experimentally constructed state $\hat{\rho}_k^\mathrm{exp}$ at time $t_j$ with initial state $\rho_k(0)=|\psi_k(0)\rangle\langle\psi_k(0)|$ and that predicted by the model $\rho_k^\mathrm{prd}(t_j)$, i.e., $\mathcal{D}_k^j=\mathcal{D}(\hat{\rho}_k^\mathrm{exp}$, $\rho_k^\mathrm{prd}(t_j))$, the values reported in the box plot in Fig.~\ref{fig:app_box} are averaged over the test data set with different initial states:
$\mathcal{D}_j = \frac{1}{4} \sum_{k\in\{1,2,3,4\}} \mathcal{D}_j^k$. 
After doing this averaging, we arrive at an array $\{\mathcal{D}_j\}_{j=1}^{j=24}$, corresponding to the trace distance, again, averaged over four different initial states in the testing data set, between the experimentally constructed states and the model's predicted states at 24 sampled time points during the evolution.

\begin{table}[ht!]
\centering	
\begin{tabular}{|c|c||c|c|}
\hline
	 $\Delta_i$ & \makecell{level of empirical support \\for model $i$} & models & \makecell{$\Delta_i$\\$=\mathrm{AIC}_i-\mathrm{AIC}_2$}\\
	\hline
 0-2  & substantial & $\mathcal{M}_0$ & 9.46 \\
4-7 & considerably less & $\mathcal{M}_1$ & 0.43 \\
>10 & essentially none & $\mathcal{M}_2$ & 0 \\
\hline
\end{tabular}
\caption{A heuristic interpretation of AIC differences $\Delta_i$ reported in Fig.~\ref{fig:1d}. The larger $\Delta_i$ is, the less plausible it is that the model $\mathcal{M}_i$ is the best model.}
\label{tab:3}
\end{table}

Finally, using the tomography data with the qubit initialized in $\rho_0(0)=\ketb{+i}{+i}$ and $\rho_2(0)=\ketb{-i}{-i}$, we reevaluate the degree of non-Markovianity for different spectator qubit states in Fig.~\ref{fig:nonMarkY} according to Eqs.~\eqref{eq:derivative} and~\eqref{eq:blp}. The non-monotonic decay in the trace distance $\mathcal{D}$ and the estimated non-Markovian measure $\mathcal{N}$ in Fig.~\ref{fig:nonMarkY} agrees well with those in Fig.~\ref{fig:nonMark}, serves as a supplementary quantitative demonstration of non-Markovian effects present in the device. The degree of non-Markovianity $\mathcal{N}^\mathrm{prd}_{\mathcal{M}_1}$ and $\mathcal{N}^\mathrm{prd}_{\mathcal{M}_2}$ calculated from the constructed PMME models agrees well with that from the experimental data $\mathcal{N}^\mathrm{exp}$, showing that the PMME models have the ability to quantitatively describe and predict the degree of non-Markovianity of the dynamics during the qubit free evolution.

\bibliography{refs.bib}

\end{document}